\documentclass[longauth]{aa}
\usepackage{graphicx}
\usepackage{txfonts}
\usepackage{url}
\begin{document}

   \title{The XXL Survey: XXVII. The \texttt{3XLSS} point source catalogue.
   \thanks{The \texttt{3XLSS} X-ray catalogue described in Tables \ref{TabBand}
   and \ref{TabMerge},
   the \texttt{3XLSSOPTN} and  \texttt{3XLSSOPTS} multiwavelength catalogues
   described in Table~\ref{TabMulti},
   the redshift catalogue (Full Table \ref{tab:redshiftCatalogue}), and the updated XMM XXL observation list
   (Full Table \ref{TabAstrometry}) are only available at CDS.
    }
   \thanks{Based on observations obtained with XMM-Newton, 
   an ESA science mission with instruments and contributions 
   directly funded by ESA Member States and NASA.}}


   \author{L. Chiappetti\inst{1}
          \and
          S. Fotopoulou\inst{2}
          \and
          C. Lidman\inst{3}
          \and                          
          L. Faccioli\inst{4,5}
          \and
          F. Pacaud\inst{6}
          \and
          A. Elyiv\inst{7,8}
          \and
          S. Paltani\inst{2}                       
          \and                          
          M. Pierre\inst{4,5}
          \and
          M. Plionis\inst{9,10}
          \and
          C. Adami\inst{11}
          \and
          S. Alis\inst{12}
          \and
          B. Altieri\inst{13}
          \and
          I. Baldry\inst{25}                 
          \and
          M. Bolzonella\inst{22}             
          \and
          A. Bongiorno\inst{14}
          \and
          M. Brown\inst{26}                  
          \and
          S. Driver\inst{27,28}              
          \and
          E. Elmer\inst{15}
          \and
          P. Franzetti\inst{1}               
          \and
          M. Grootes\inst{29}                
          \and
          V. Guglielmo\inst{16,11}
          \and
          A. Iovino\inst{17}
          \and
          E. Koulouridis\inst{4,5}
          \and
          J.P. Lef\`evre\inst{4,5}
          \and
          J. Liske\inst{30}                  
          \and
          S. Maurogordato\inst{18}
          \and
          O. Melnyk\inst{8,19}
          \and
          M. Owers\inst{3,20}                
          \and
          B. Poggianti\inst{16}
          \and
          M. Polletta\inst{1,31}             
          \and
          E. Pompei\inst{21}
          \and
          T. Ponman\inst{32}                 
          \and
          A. Robotham\inst{27,28}            
          \and
          T. Sadibekova\inst{4,5}
          \and
          R. Tuffs\inst{29}                 
          \and
          I. Valtchanov\inst{13}
          \and
          C. Vignali\inst{7,22}
          \and
          G. Wagner\inst{23,24}
          }

   \institute{INAF, IASF Milano, via Bassini 15, I-20133 Milano, Italy;
              \email{lucio@lambrate.inaf.it}
         \and                                                                                                    
              Department of Astronomy, University of Geneva, ch. d'\'Ecogia 16, CH-1290 Versoix, Switzerland
         \and                                                                                                    
              Australian Astronomical Observatory, North Ryde, NSW 2113, Australia
         \and                                                                                                    
              IRFU, CEA, Universit\'e Paris-Saclay, F-91191 Gif sur Yvette, France
         \and                                                                                                    
              Universit\'e Paris Diderot, AIM,  Sorbonne Paris Cit\'e, CEA, CNRS,  F-91191 Gif sur Yvette,  France
         \and                                                                                                    
              Argelander Institut f\"ur Astronomie, Universit\"at Bonn, D-53121 Bonn, Germany
         \and                                                                                                    
              Dipartimento di Fisica e Astronomia, Universit\`a di Bologna, via Gobetti 93/2, I-40129 Bologna, Italy
         \and
              Main Astronomical Observatory, Academy of Sciences of Ukraine, 27 Akademika Zabolotnoho St., 03680 Kyiv, Ukraine
         \and
         National Observatory of Athens, Lofos Nymfon, GR-11810 Athens, Greece                                     
         \and                                                                                                    
              Aristotle University of Thessaloniki, Physics Department, Thessaloniki, GR-54124, Greece
         \and                                                                                                    
              Aix Marseille Univ., CNRS, LAM, Laboratoire d'Astrophysique de Marseille, F-13388 Marseille, France
         \and
         Department of Astronomy and Space Sciences, Faculty of Science, Istanbul University, 34119 Istanbul, Turkey            
         \and
         European Space Astronomy Centre (ESA/ESAC), Villanueva de la Ca\~nada, E-28692 Madrid, Spain         
         \and
         INAF, Osservatorio Astronomico di Roma, via Frascati 33, I-00078 Monte Porzio Catone, Italy        
         \and                                                                                                    
              School of Physics and Astronomy, University of Nottingham, Nottingham NG7 2RD, United Kingdom
         \and                                                                                                    
              INAF, Osservatorio Astronomico di Padova, vicolo dell'Osservatorio 5, I-35122 Padova, Italy
         \and
         INAF, Osservatorio Astronomico di Brera, Via Brera 28, I-20122 Milano, Italy                       
         \and
         Laboratoire Lagrange, UMR 7293, Universit\'e de Nice Sophia Antipolis, CNRS,                       
         Observatoire de la C\^ote dAzur, F-06304 Nice, France
         \and                                                                                               
         Department of Physics, University of Zagreb, Bijeni\v{c}ka cesta 32, 10000  Zagreb, Croatia
         \and
         Macquarie University, NSW 2109, Australia                                                          
         \and
         European Southern Observatory, Alonso de Cordova 3107, Vitacura, 19001 Casilla, Santiago 19, Chile 
         \and
         INAF, Osservatorio Astronomico di Bologna, Via Gobetti 93/3 I-40129 Bologna, Italy                 
         \and                                                                                                    
              Department of Physics, University of Oxford, Oxford OX1 3PU, United Kingdom
         \and                                                                                                    
              Merton College, Oxford OX1 4JD, United Kingdom
         \and                                                                                                    
              Astrophysics Research Institute, Liverpool John Moores University, Liverpool L3 5RF, United Kingdom
         \and
         Monash University, Victoria 3800, Australia                                                        
         \and
         ICRAR, 1 Turner Avenue, Technology Park, Bentley, Western Australia                                
         \and
         University of St Andrews, St Andrews, KY16 9AJ, Scotland, United Kingdom                           
         \and
         Max Planck Institut f\"ur Kernphysik, Saupfercheckweg 1, D-69117 Heidelberg, Germany               
         \and
         Hamburger Sternwarte, Universit{\"at} Hamburg, Gojenbergsweg 112, D-21029 Hamburg, Germany                                  
         \and
         IRAP, Universit\'e de Toulouse, CNRS, UPS, F-31400 Toulouse, France                                
         \and
         School of Physics and Astronomy, University of Birmingham, B152TT Birmingham,  United Kingdom      
             }
                                             
   \date{Received ; accepted }

 

\abstract{
We present the version of the point source catalogue of  the XXL Survey that was 
used, in part, in the first series of XXL papers. 
In this paper we release, in our database in Milan and at CDS:
(i) the X-ray source catalogue with 26\,056 objects in two areas of 25 deg$^2$
with a flux limit (at 3$\sigma$) of
$\sim 10^{-15}$ erg s$^{-1}$ cm$^{-2}$ in [0.5-2] keV, and
$\sim$ 3 $10^{-15}$ erg s$^{-1}$ cm$^{-2}$ in [2-10] keV,
yielding a 90\% completeness limit of
5.8 $10^{-15}$ and 3.8 $10^{-14}$ respectively
;
(ii) the associated multiwavelength catalogues with candidate counterparts
of the X-ray sources in the infrared, near-infrared, optical, and ultraviolet (plus spectroscopic redshift when available);
and 
(iii) a catalogue of spectroscopic redshifts recently obtained in the southern XXL area.
We also present  the basic properties of the X-ray point sources and their counterparts.
Other catalogues described in the second series of XXL papers will be released contextually,
and will constitute the second XXL data release.
}

   \maketitle

%
   \defcitealias{2016A&A...592A...1P}{XXL~Paper~I}
   \defcitealias{2016A&A...592A...2P}{XXL~Paper~II}
   \defcitealias{2016A&A...592A...5F}{XXL~Paper~VI}
   \defcitealias{2016A&A...592A...8B}{XXL~Paper~IX}
   \defcitealias{2016A&A...592A..10S}{XXL~Paper~XI}
   \defcitealias{2016PASA...33....1L}{XXL~Paper~XIV}
   \defcitealias{2017A&A..paperXVIII}{XXL~Paper~XVIII}
   \defcitealias{2017A&A.....paperXX}{XXL~Paper~XX}
   \defcitealias{2017A&A....paperXXI}{XXL~Paper~XXI}
   \defcitealias{2017A&A...paperXXII}{XXL~Paper~XXII}
   \defcitealias{2017A&A...paperXXIV}{XXL~Paper~XXIV}
   \defcitealias{2017A&A...paperXXIX}{XXL~Paper~XXIX}
   \defcitealias{2018A&A...paperXVI}{XXL~Paper~XVI}
   \defcitealias{2018A&A...paperXXVIII}{XXL~Paper~XXVIII}
%

\section{Introduction\label{SecIntro}}

   The XXL Survey is an X-ray survey carried out by the XMM-Newton satellite, covering
   two 25 deg$^2$ areas 
   called XXL-N and XXL-S,
   complemented by observations at multiple wavelengths. 
   Initial results from the survey 
   were
   published in a dedicated issue of
   Astronomy \& Astrophysics (volume 592).
   For the scientific motivations
   of the survey,
   the characteristics of the observing programme,  and the multiwavelength
   follow-up programmes, refer to \citet[hereafter \citetalias{2016A&A...592A...1P}]{2016A&A...592A...1P}.

The first XXL data release (hereafter DR1), which is associated with the first series of papers,
   included the catalogue of the 100 brightest galaxy clusters, published in
   \citet[hereafter \citetalias{2016A&A...592A...2P}]{2016A&A...592A...2P},
   and the 
   catalogue
   of the 1000 brightest point-like sources, published in
   \citet[hereafter \citetalias{2016A&A...592A...5F}]{2016A&A...592A...5F}.
   A second 
   release (DR2) 
   will occur
   jointly with the publication of the present paper.
   A list of the contents of
   DR1 and DR2
      (inclusive of references) is tabulated in  section~\ref{SecDB}.

   We present here the
   catalogue of all the sources 
   used for DR1 and DR2, from which the above
   DR1
   catalogues
   were drawn
  (as well as a recent study on the environment and clustering of AGN,
  \citealt{2017A&A....paperXXI},
  also known as \citetalias{2017A&A....paperXXI}).
   After DR2,
   the XXL collaboration plans 
   to reprocess
   the data with an improved pipeline,
   described in
   \citet[hereafter \citetalias{2017A&A...paperXXIV}]{2017A&A...paperXXIV}.

   The plan of the paper is as follows. Section~\ref{SecXray} presents the X-ray data,
   the X-ray processing pipeline (\ref{SecPipe}), and 
   basic properties such as sky coverage, $log N-log S$, and flux distribution (\ref{SecInfo}).
   Section~\ref{SecMW} presents the multiwavelength data, 
   describes the counterpart association (\ref{SecIdent}),
   provides
   some statistics (\ref{SecMWStat}),
   presents the spectroscopic redshifts (\ref{SecOurSpe}), and 
   describes the additional spectra obtained with the AAOmega
   spectrograph in 2016
    (\ref{SecLidman}), supplementing those published in
   \citet[hereafter \citetalias{2016PASA...33....1L}]{2016PASA...33....1L}.
   Section~\ref{SecDB} presents the Master Catalogue database site, 
   the X-ray catalogue tables (\ref{SecXrayTab}) with associated data products (\ref{SecXrayDP}),
and
   the multi-$\lambda$ catalogue tables (\ref{SecMWTab}) with associated data products (\ref{SecMWDP}).
   Section~\ref{SecEnd} summarises the work.    Appendix~\ref{SecCompLSS} compares the present catalogue with the previous XMM-LSS catalogue
   \citep{2013MNRAS.429.1652C}.

\section{X-ray material\label{SecXray}}

  The X-ray (XMM-Newton) observations of the XXL survey were obtained over several years in two sky regions as
  a collection of uniformly spaced contiguous pointings with an exposure time of at least 10 ks (complemented by a few GO pointings from
  the archives, and by pointings from earlier surveys, sometimes deeper).
  The list of the 622 XMM pointings (294 in the northern area
  at $\delta\sim -4.5\degr$,
  hereafter XXL-N; and 328 in the southern area
  at $\delta\sim -55\degr$,
  hereafter XXL-S) 
  is given in Appendix B of
  \citetalias{2016A&A...592A...1P} 
  (which also provides details on the observing strategy),
  and is available at the
  Centre de Donn\'ees astronomiques de Strasbourg (CDS) and in our database
  (see section~\ref{SecDB}
  and Appendix~\ref{SecNewApp}).

\subsection{The X-ray pipeline\label{SecPipe}}

  We use version 3.3.2 of
  the \textsc{Xamin} X-ray pipeline \citet{2006MNRAS.372..578P}
  to process the XMM X-ray data. Earlier versions of the pipeline were used
  to process data
  for the production of the XMM-LSS survey catalogues, which can be
  considered the precursors of the present catalogue on a smaller area (see Table~\ref{TabCat}).
  For this reason we 
  prefix all sources in the current catalogue with
  \texttt{3XLSS}.

  With respect to the newer pipeline reference paper,
  \citepalias{2017A&A...paperXXIV},
  all catalogues produced to date, including the present one, use 
  pipeline versions collectively grouped as the
  basic version
  called \texttt{XAminP06} in paper XXIV; in other words,  detections are done on each pointing
  separately.

  Version 3.3.2 of the pipeline differs from earlier versions in relatively minor details, including
  an optimisation 
  of the code
  for the detection of point
  sources, especially bright ones, 
  with
  a fix for some numeric problems.
  They will be retained
  in the newer \texttt{XAminF18} pipeline presented in \citetalias{2017A&A...paperXXIV}.

  \begin{table}
   \caption{XMM-LSS and XXL X-ray catalogues}
   \label{TabCat}
   \centering
   \begin{tabular}{rllrrc}
    \hline\hline
    Catalogue       & \textsc{Xamin} && Area      & N. of   & Ref. \\
    acronym       & version        && (deg$^2$) & sources &      \\
    \hline
    \texttt{XLSS}  & 3.1    &&           5.5    &  3\,385   & 1 \\
    \texttt{2XLSS} & 3.2    &&           11.1   &  5\,572   & 2 \\
    \texttt{2XLSSd} & 3.2   &&           11.1   &  6\,721   & 2 \\
    \texttt{3XLSS} & 3.3.2  & XXL-N &      25   &  14\,168  & 3 \\
                   &        & XXL-S &      25   &  11\,888  &   \\
    \hline
   \end{tabular}
   \tablebib{
    (1)~\citet{2007MNRAS.382..279P};
    (2)~\citet{2013MNRAS.429.1652C};
    (3)~this paper.
   }
  \end{table}

  As described in section 2.1 of \citetalias{2017A&A...paperXXIV},
  the basic pipeline starts from event lists filtered for soft proton flares,
  produced by standard \textsc{sas} tasks, to generate wavelet images and 
  ends with
  a \textsc{sextractor} 
  source list. \textsc{Xamin} characterises the sources with a maximum likelihood
  fit with both a pointlike and an extended ($\beta$-profile) model, and provides
   a \textsc{fits} file for each pointing with basic parameters separately for the
  soft [0.5-2] keV, or B, band and for the hard [2-10] keV, or CD, band.

\begin{figure*}
 \includegraphics[width=17.0cm]{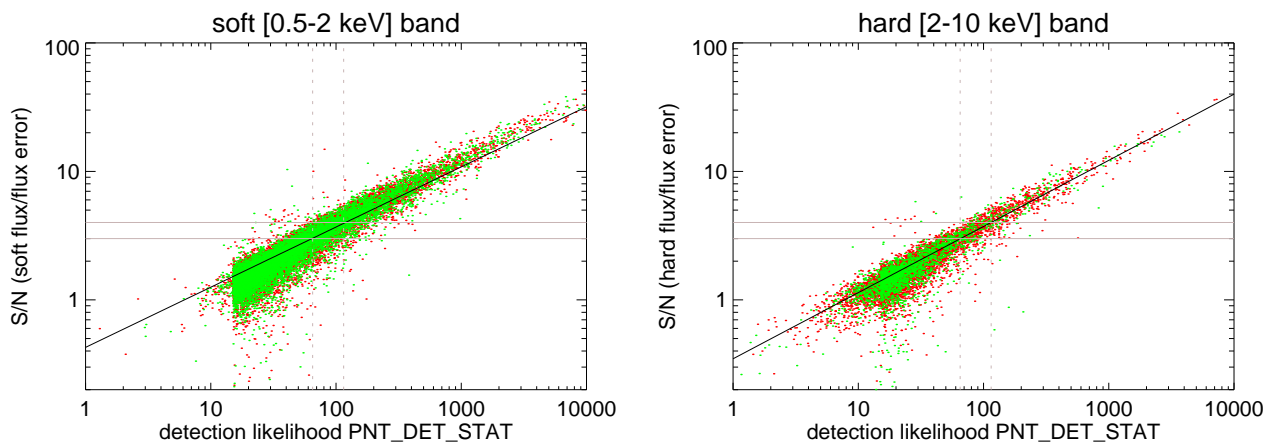}
 \caption[dummy caption]{
 Cross-calibration of the S/N  (flux divided by computed flux error) vs the
 detection 
 statistic
  \textsc{pnt\_det\_stat} for the soft (left) and hard (right)
 X-ray bands. The red and green dots correspond respectively to point-like sources in XXL-N and XXL-S.
 The solid thick line is 
 a linear fit in log-log space to the S/N  averaged
 in pseudo-logarithmic\footnotemark{} 
 likelihood bins.
 The two pale grey  horizontal lines are fiducial marks for the $3\sigma$ and
 $4\sigma$ levels. The equivalent dotted vertical lines are for \textsc{pnt\_det\_stat} of 65 and
 115.
  }
 \label{FigSigma}
\end{figure*}

The parameters produced by
  \textsc{Xamin} 
  (which can be summarised as exposure times, statistics, raw source positions,
  source and background counts, and source count rates)
  are listed in detail in Table 1 of \citetalias{2017A&A...paperXXIV}
  and Table 2 of \citet{2006MNRAS.372..578P}, and 
  are
  flagged in  Col. 4 (`X')
  of Tables \ref{TabBand} and \ref{TabMerge}
  of this paper.
  Other parameters are computed a posteriori
  during database ingestion
  on individual bands, as in \citet{2013MNRAS.429.1652C}. These  parameters include the following:
  \begin{itemize}

  \item The application of the C1/C2 recipe to classify extended sources 
  \citep{2006MNRAS.372..578P},
  which are characterised by 
\smallskip \\
the C1 recipe for
  a sample of clusters uncontaminated by point-like sources,
   \\
  \ \ \ \textsc{ext $>$ 5\arcsec and ext\_stat $>$ 33 and ext\_det\_stat $>$ 32}, and
\smallskip \\
the C2 recipe for
  clusters allowing a 50\% contamination by misclassified point sources,
 \\
  \ \ \ \textsc{ext $>$ 5\arcsec and ext\_stat $>$ 15};
\smallskip 

  Although validated extensively only for the soft band, the recipe is nominally applied also to the hard band.
  A detection 
  that
  satisfies either C1 or C2 in a band is flagged extended in that band (and by
  definition not spurious).
  The source 
  extent,
  \textsc{ext}, the
  so-called
  detection likelihoods\footnote{
  The detection statistic used by \textsc{Xamin}, although  customarily but
  improperly referred to as  likelihood, is 
  a modified Cash statistic, and the extension statistic \textsc{ext\_stat} is
  function of the difference of the detection statistics as pointlike
  and as extended. The detection statistic is a linear function of the logarithm of the true likelihood. The 
  formulae used for computation are reported in section 2.3.1 of \citet{2006MNRAS.372..578P}.
  },
   \textsc{ext\_det\_stat} (fit \textit{as extended}) and 
   \textsc{ext\_pnt\_stat} (fit \textit{as pointlike}),
  and the likelihood of being extended, \textsc{ext\_stat}, are \textsc{Xamin} parameters for which we follow the notation used in
  \citetalias{2017A&A...paperXXIV}. The correspondence to database column names is reported in Table~\ref{TabBand}.

  \item The application of the P1 recipe,
{which aims to  define a sample of point sources with a high degree of
 purity and which is complete down to the lowest count rate possible
}
  \citepalias{2017A&A...paperXXIV},
  used for the first time in the \texttt{3XLSS} catalogue,
  characterised  by \\
  \ \ \ \textsc{pnt\_det\_stat $>$ 30 and (ext $<$ 3\arcsec  or ext\_stat=0)}.

  Stricly speaking, the sources which are not C1, C2, or P1 are to be considered
  undefined (i.e. there are not enough photons to unambiguously characterise them).
  The undefined sources are flagged as pointlike in the database.
  Those
  with \textsc{pnt\_det\_stat $<$ 15} are considered spurious in the band (i.e. not considered
  for inclusion in the catalogue). 

  \item Flux conversion. A conventional mean flux $(FLUX_{MOS}+FLUX_{pn})/2$ in both
  bands is computed from count rates via the usual conversion factors CF 
  listed in
  Table~\ref{TabCF} with a fixed spectral model 
  ($\Gamma=1.7$,  N$_H=2.6\times10^{20}$ cm$^{-2}$).

  \item Error computation. \textsc{Xamin} currently does not provide errors on
  count rates, fluxes, or position. A positional error is computed 
  as a function of count rate and off-axis angle 
  (e.g. Table 4 of \citealt{2013MNRAS.429.1652C}).
  Flux errors are
  computed by calculating the Poissonian error on gross counts\footnote{
  Gross counts are \textit{reconstructed} 
  by
  adding net counts and background counts in the extraction area,
  which are parameters 
  produced by \textsc{Xamin}.
  The Gehrels formula on gross counts was
  used, e.g. in XMDS \citep{2005A&A...439..413C}
  and HELLAS2XMM \citep{2002ApJ...564..190B}.
  }
  according to the formula of \citet{1986ApJ...303..336G}, assuming that the fractional error on
  rates is the same as that on counts, and propagating them through the CF formula.
In Fig.~\ref{FigSigma} we 
compare
the S/N  using the computed flux errors
to
the \textsc{Xamin} detection 
statistic.
The customary $3\sigma$ and $4\sigma$ levels seem
to occur for likelihood levels $\sim$ 65 and 
115,
higher than those 
found in the comparison
of \textsc{Xamin} 3.2
with the XMDS pipeline given in \citet{2013MNRAS.429.1652C}, which only deals  with some of the longest exposures
in XXL, 
and uses a different analysis software.

  \end{itemize}

Additional
  steps are performed after the ingestion of the band tables in the database
  for the preparation of the catalogues. 
  These steps are described next.

  \begin{table}
   \caption{Conversion factor CF from count rate to flux in units of
   10$^{12}$ erg s$^{-1}$ cm$^{-2}$ for a rate of one count s$^{-1}$ . A photon-index power law
   with $\Gamma=1.7$ and a mean N$_H$ value of $2.6\times10^{20}$ cm$^{-2}$ are assumed.
   The two MOS cameras are assumed to be identical.}
   \label{TabCF}
   \centering
   \begin{tabular}{lrr}
    \hline\hline
    EPIC camera   & soft (B) band  & hard (CD) band \\
    \hline
    MOS           & 5.0            & 23.0 \\
    pn            & 1.5            &  7.9 \\
    \hline
   \end{tabular}
  \end{table}

  \subsubsection{Band merging \label{SecBandMerge}}
  Band merging of detections in the soft and hard bands
  occurs in a 10 arcsec radius 
  following the
  procedure used for
  XMM-LSS and described in detail in \citet{2013MNRAS.429.1652C}.
  A merged source can be classified 
(using a two-letter code, \texttt{P} for pointlike, \texttt{E} for extended,
\texttt{-} for undetected, respectively in the soft and hard bands) 
  as
  extended (and by definition non-spurious) if it is 
  extended in both
  bands (\texttt{EE}) or in the single one where 
  it
  is detected (\texttt{E-} and \texttt{-E}),
  or
  if it is extended in the soft band (\texttt{EP}).
  It is classified as pointlike in all the other cases (\texttt{PP}, \texttt{P-}, \texttt{-P}, and
  \texttt{PE}).
  \\
  Band merging acts on all detections in all pointings, including those 
  considered as spurious 
  (i.e. detection statistic
  \textsc{det\_stat$<$15} in both bands), 
  which 
  are not included
  in the released catalogue. We have
  26\,555 merged detections in XXL-N, of which 17\,398
  are
  non-spurious.
  The respective numbers for XXL-S are
  27\,173 and 18\,145.
  \\ 
  A band-merged detection in two bands 
  that
  is spurious in one band (flagged by
  the boolean flags
  \texttt{Bspurious=1} or \texttt{CDspurious=1}) 
  still
  provides some 
  usable information
  (rate, flux, etc.) and is 
  retained
  in the catalogue.
 \\
  For detections in two bands, 
  the band in which the detection likelihood of the source is
  the highest 
  is the band from which the coordinates are taken. 
  If we compare the 
  separation
  between B and CD positions
  (column \texttt{Xmaxdist} in the database)
  with the positional uncertainties in the two bands added in quadrature ($\sigma$)
  we 
  find
  that
  \texttt{Xmaxdist}$<\sigma$ in 33\% of the cases, within $2\sigma$ in 71\%
  of the cases
  and within $3\sigma$ in 77\%,
  which looks reasonable.
 \\
  It is possible to have an ambiguous band merging
  case
  when a detection
  in a band happens to be associated with two different objects in the
  other band (i.e. it gives rise to two entries in the merged table).
  If the two entries both have \texttt{Xmaxdist}$<6\arcsec$ or
  \texttt{Xmaxdist}$>6\arcsec$, they are 
  both retained
  as intrinsically ambiguous.
  If one is below 6\arcsec and one above, the lower-distance entry remains a 
  merged two-band detection, while the other is `divorced' as a hard-only or 
  soft-only source.
 \\
  The naming of ambiguous sources is described in section~\ref{SecXrayTab}.
  \\ 
  The number of 
  such
  ambiguous 
  cases in the final catalogue (i.e.
  after the next step of overlap handling) is very limited:
  74 
  out of
  14\,168
  sources
  in XXL-N
  and 69 
  out of
  11\,888
  sources
  in XXL-S, i.e. approximately 0.5\% of the total.
  For comparison, 0.7\% of the sources in the \texttt{2XLSS} catalogue \citep{2013MNRAS.429.1652C}
  were ambiguous.

\footnotetext{
The pseudo-logarithmic binning is a spacing of 1 in
\textsc{pnt\_det\_stat} up to 100, then  5 up to 1000,  50 up to 10000, and  500
above 10000.}

  \subsubsection{Pointing overlap removal}
  Accounting for duplicate detections in pointing overlap regions
  (`overlap removal')
  is the final stage of catalogue generation.
  Consistently with the selection criteria defined in \citet{2006MNRAS.372..578P},
  aimed at the best balance between contamination and completeness, and  already used
  in \citet{2007MNRAS.382..279P} and \citet{2013MNRAS.429.1652C},
  detections with \textsc{det\_stat$<$15} are discarded at this stage and only non-spurious sources
  are brought forward as catalogue sources
  (for an object to be kept a non-spurious detection in one band is sufficient).

  Detections occurring in a single pointing are always kept irrespective of the pointing quality; instead,
  in all cases where two or more detections occur 
  within 10\arcsec~of one another
  in different pointings, the same 
  procedure used for XMM-LSS and
  described in \citet{2013MNRAS.429.1652C} is applied, i.e. if one detection is in a
  pointing with a better bad-field flag (column \texttt{Xbadfield} in the database),
  then it
   is kept in preference to any other. 
  Otherwise if \texttt{Xbadfield} is the same, the source closest to the 
  XMM pointing
  centre
  (columns \texttt{Boffaxis} and \texttt{CDoffaxis} in the database) is kept. There is,
  however, a difference with the XMM-LSS case: 
  \texttt{Xbadfield} is not a 
  boolean 
  flag, but 
  can have
  a range of values:

  \begin{itemize}
  \item 0 : pointing belongs to XXL, has good quality, and is the single or best
            pointing of a sequence of repeats;
  \item 1 : pointing does not belong to XXL (archival or deep followup pointing) and has good quality;
  \item 2 :  another XXL pointing 
  in a sequence of repeated pointings (see below) that has good quality;
  \item 3 : quality is bad (which may occur if the exposure is too
  shallow,
  or background is too high, or both).
  \end{itemize}
  We note that 94\% of the sources have \texttt{Xbadfield=0} and only 2\% have
  \texttt{Xbadfield=3}.

  One should remember that a given pointing might have been 
  re-observed 
  (repeats) if 
  previous observation(s)
  did not achieve the scheduled exposure. Repeats are
  characterised by a \texttt{XFieldName} of the form \texttt{XXL{\it smmm-ppc}} with a different letter
  \texttt{{\it c}=a,b,c,\ldots} in the rightmost position.
  In some cases it has been possible to combine
  the event files of 
  repeats (called z-pointings because the last letter 
  in
  the \texttt{XFieldName} is conventionally \texttt{z}, e.g. \texttt{XXLn000-04z}):
  z-pointings are by definition considered
  superior to the individual repeats in the sequence.

\begin{figure*}
 \includegraphics[width=17.0cm]{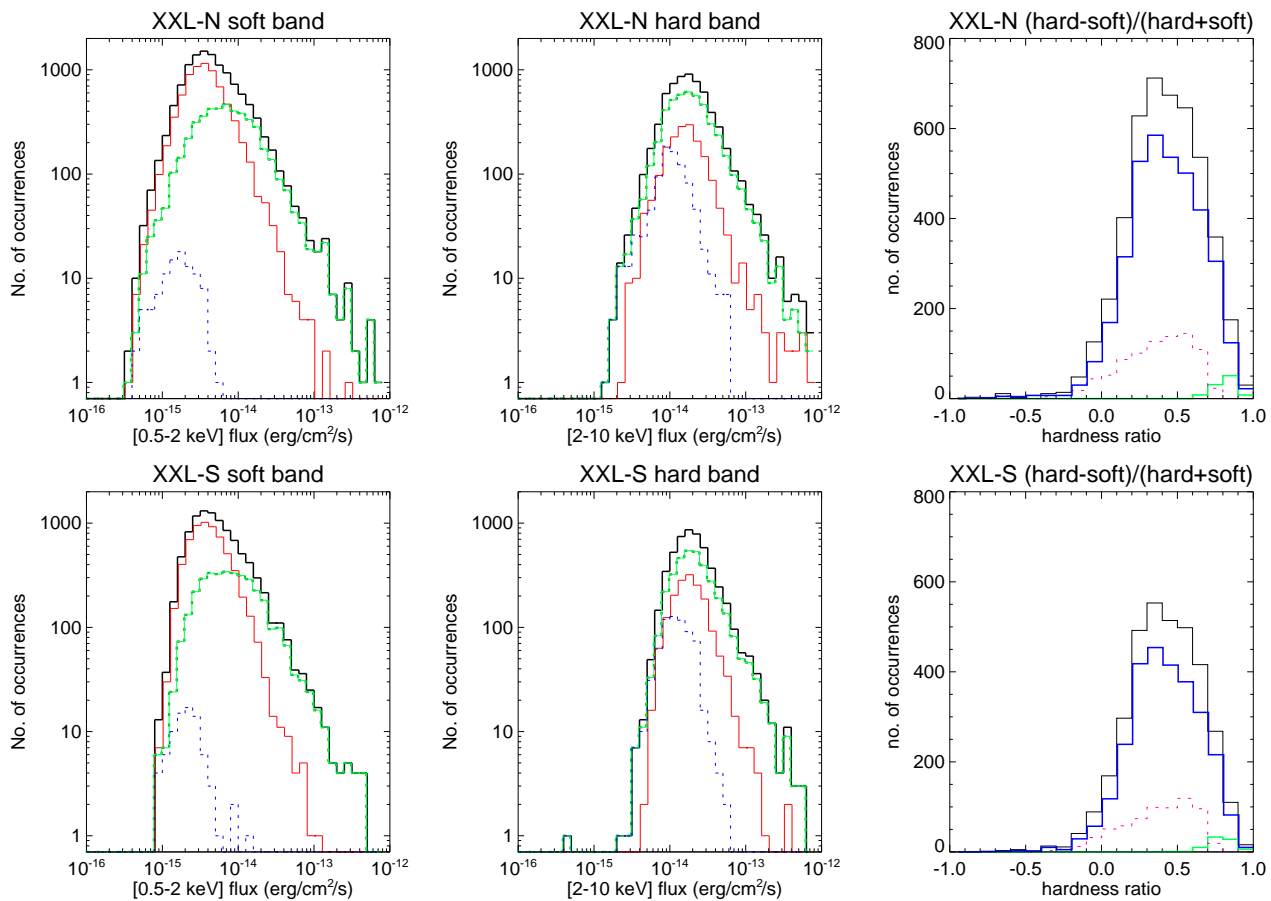}
 \caption{
 Distribution of the X-ray flux and hardness ratio (HR) for XXL-N (top panels) and XXL-S (bottom panels).
 The left column 
 shows
 the soft band, 
 the centre column the hard band, 
 and the right
 column the flux hardness ratio (customarily defined as [hard-soft]/[hard+soft]), computed for sources detected
 in both bands.
 The black histograms in the four leftmost plots correspond
 to all sources detected \textit{in
 each
 band};
 the thin red histograms to sources detected \textit{only in 
 one
 band} (hence by definition not spurious in it);
 the green dashed histograms to sources also detected  in the other band (the sum of the green and red histograms
 corresponds to the total); the subset of the sources detected in both bands but nominally spurious in the
 band of relevance are shown by the dotted blue histogram.
 In the HR distribution, the thin black histogram corresponds to all sources detected in both bands under any condition,
 which in turn fall into the following three categories:
 the thick blue histogram 
 represents sources that
 are non-spurious in both bands;
 the dashed magenta one 
 represents those that are
 spurious in the hard band;
and  the tiny thick green one at the extreme right 
 those that are
 spurious in the soft band.
 Some histograms are slightly displaced 
 along the
 x-axis for clarity.
  }
 \label{FigFlux}
\end{figure*}

  \subsubsection{Astrometric correction \label{SecAstroCorr}}
 Following the procedure used for XMM-LSS, an astrometric correction was applied to
 all sources using the \textsc{sas} task \textsc{eposcorr},
 applying to all positions in a pointing a global rigid shift (no rotation was applied).
  The offsets
  were computed using reference optical catalogues. For XXL-N the
  CFHT Legacy Survey
  T007 version
  from Vizier (catalogue II/317, \citealp{2012yCat.2317....0H})
  was used,
  except for one pointing
  where the USNO A2.0 catalogue was used
  and 16 pointings for which no correction could be computed (bad pointings,
  with few X-ray detections, which are usually ignored by the overlap removal procedure
  in favour of better repeats of the same pointing). For XXL-S the reference catalogue
  was from the Blanco Cosmology Survey \citep[BCS, ][]{2009ApJ...698.1221M}, except for 11 pointings for which no correction
  could be computed for the same reasons 
  noted above.
  The 
  offsets per pointing, available in web pages reachable from
  the `table help pages' in the database,
  have been added as new columns in
  the pointing list table, already published in \citetalias{2016A&A...592A...1P};
  the updated table is presented in Appendix~\ref{SecNewApp}.

\subsection{Basic properties\label{SecInfo}}

 Table~\ref{TabStat} provides 
 the number of
 pointlike and extended sources in the catalogue, 
 split by
 the C1, C2, and P1 recipes
 described above.

In Fig.~\ref{FigFlux} we 
show the X-ray flux distributions for both the soft and hard bands, and
the hardness ratio 
(customarily defined as [hard-soft]/[hard+soft])
distribution
(for sources detected in both bands) for XXL-N and XXL-S. 

The relationship between flux and detection 
statistic
 \textsc{pnt\_det\_stat}
  was reported in Fig.~4 of \citetalias{2016A&A...592A...1P}.
  Therefore, we felt 
  it unnecessary 
  to repeat the information here.

  \begin{table}
   \caption{Source numbers for the X-ray catalogues.
   }
   \label{TabStat}
   \centering
   \begin{tabular}{lrr}
    \hline\hline
    N. of sources                       & XXL-N          & XXL-S   \\
    \hline
    total                               & 14\,168          & 11\,888   \\
    \hline
    ~pointlike                          & 13\,770          & 11\,413   \\
    ~~     \ldots and P1                &  7\,246          &  5\,565   \\
    ~~\texttt{PP} (detected in 2 bands) &  4\,597 &  3\,479 \\
    ~~~     \ldots and P1               &  3\,917 &  2\,907 \\
    ~~\texttt{PE} (detected in 2 bands) &      14 &      21 \\
    ~~~     \ldots and P1               &      10 &      19 \\
    ~~\texttt{P-} (only soft)           &  7\,389 &  6\,247 \\
    ~~~     \ldots and P1               &  3\,085 &  2\,451 \\
    ~~\texttt{-P} (only hard)           &  1\,770 &  1\,666 \\
    ~~~     \ldots and P1               &     234 &     188 \\
    \hline
    ~extended for \textsc{Xamin}        &   398 &   475 \\
    ~~soft C1 (\texttt{Bc1c2=1} in database)  &   136 &   118 \\
    ~~soft C2 (\texttt{Bc1c2=2} in database)  &   188 &   208 \\
    ~~hard-only C1 or C2 (\texttt{-E})  &    74 &   149 \\            
    ~~XLSSC clusters\tablefootmark{a}   &   186 &   142 \\            
    \hline
   \end{tabular}
\tablefoot{\tablefoottext{a}{
All of these are present in the 365 cluster catalogue of \citetalias{2017A&A.....paperXX},
but two southern objects (XLSSC~613 and 630) flagged `tentative';
17 clusters from \citetalias{2017A&A.....paperXX} (all but one in Table~G.1 of such paper) are not listed in
our pointlike catalogues because the original X-ray detection was flagged point-like \textit{and spurious};
22 clusters in Table~G.2 of \citetalias{2017A&A.....paperXX} have no XLSSC number assigned, but could be
matched to our pointlike catalogues using the \texttt{Xcatname} database column.
}
}
  \end{table}

\begin{figure}
 \includegraphics[width=8.2cm]{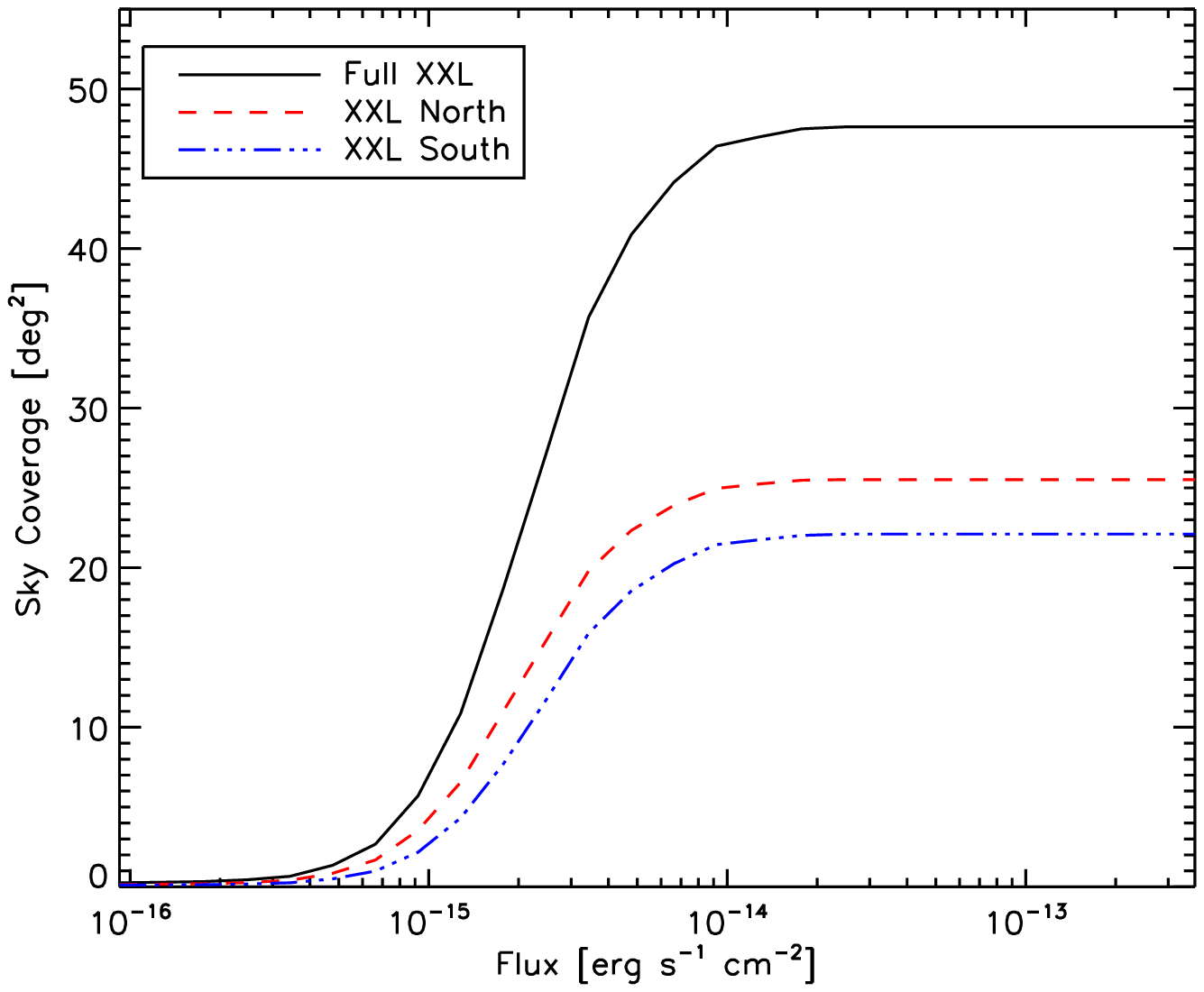}
 \includegraphics[width=8.2cm]{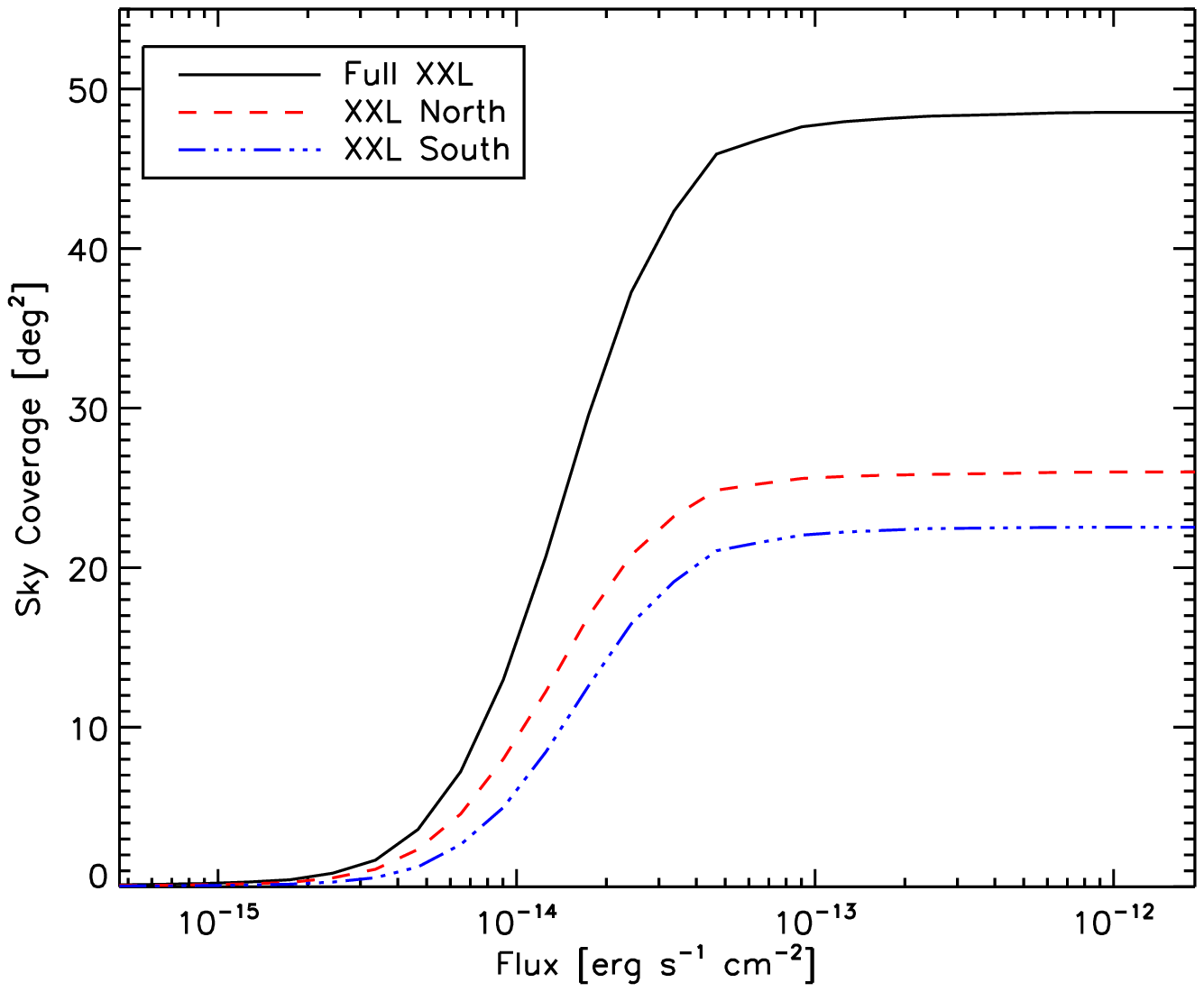}
 \caption{Sky coverage plots in the soft (top panel) and hard (bottom panel) bands. 
          The dashed and dash-dotted lines correspond to XXL-N and XXL-S, and the solid
          line to the combination of the two.
  }
 \label{FigSkyCov}
\end{figure}

The cross-calibration of the S/N  using the computed flux errors versus the \textsc{Xamin} detection statistic
has been reported above in Fig.~\ref{FigSigma}.

  The photometric accuracy of the present pipeline is presented in Fig.~4 of \citet{2013MNRAS.429.1652C},
  to which we refer
  the reader
  since it is unaffected by the changes in the pipeline version.

\subsubsection{Sky coverage\label{SecInfo1}}

 The sky coverage for 
 XXL-N and XXL-S
 is shown in Fig.~\ref{FigSkyCov}.
 Removing bad fields, i.e. keeping only
 (\texttt{Xbadfield}$<3$),
 the sky coverage is
 95\% of that shown.
 The process of removing duplicates
 keeps sources in 273 northern and 321 southern pointings,
 of which respectively 242 and 288 are good. However, since the removal operates on individual
 sources, 98\% of the catalogue  entries are in good pointings (only 313 XXL-N and 261 XXL-S sources
 are in bad 
 pointings
 ).

 The 90\% of the coverage is achieved for a flux of
5.8 $10^{-15}$ erg s$^{-1}$ cm$^{-2}$ in the soft band, and of
3.8 $10^{-14}$ erg s$^{-1}$ cm$^{-2}$ in the hard band.

\subsubsection{Source counts\label{SecInfo2}}

\begin{figure}
 \includegraphics[          width=8.2cm]{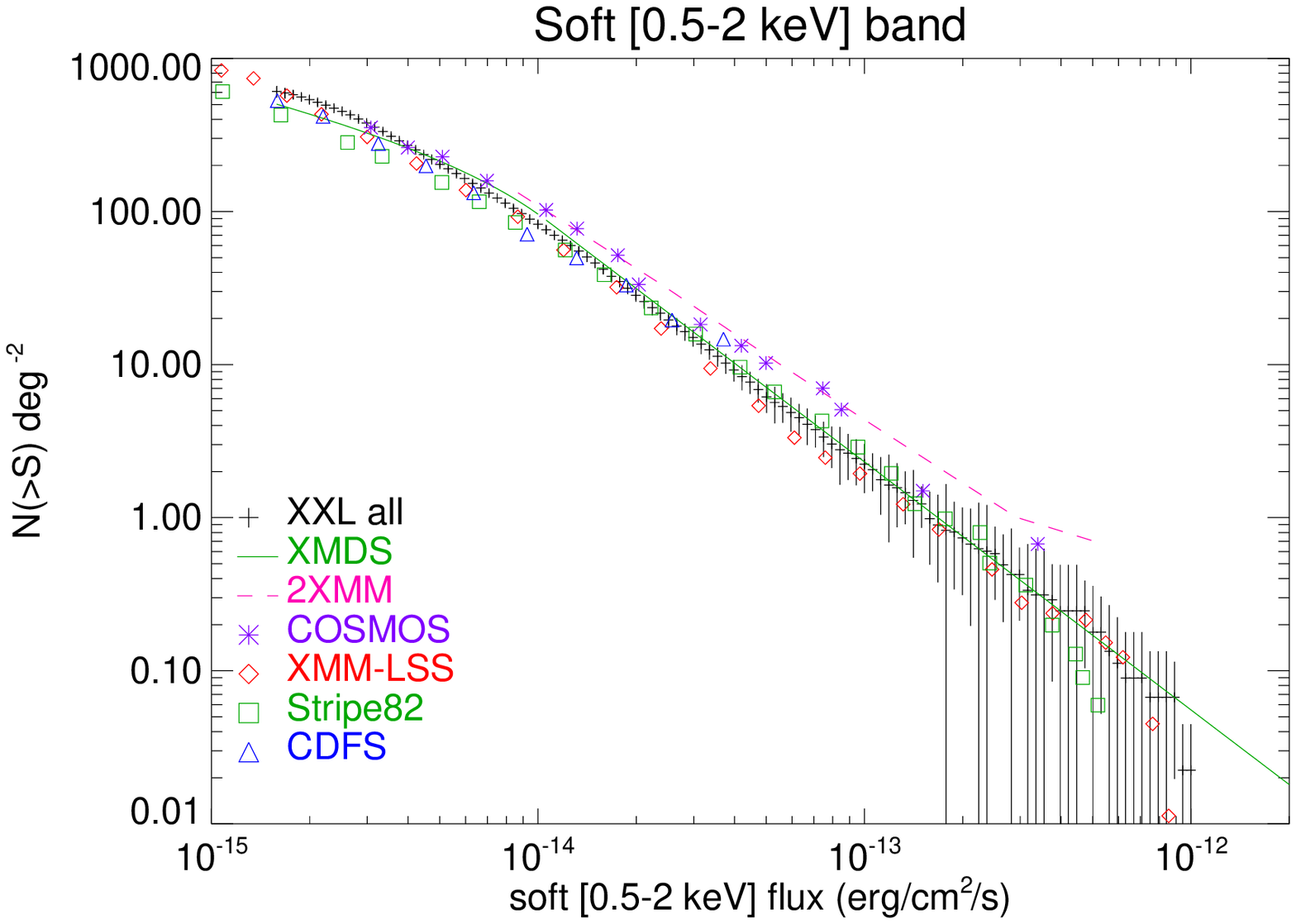}
 \includegraphics[          width=8.2cm]{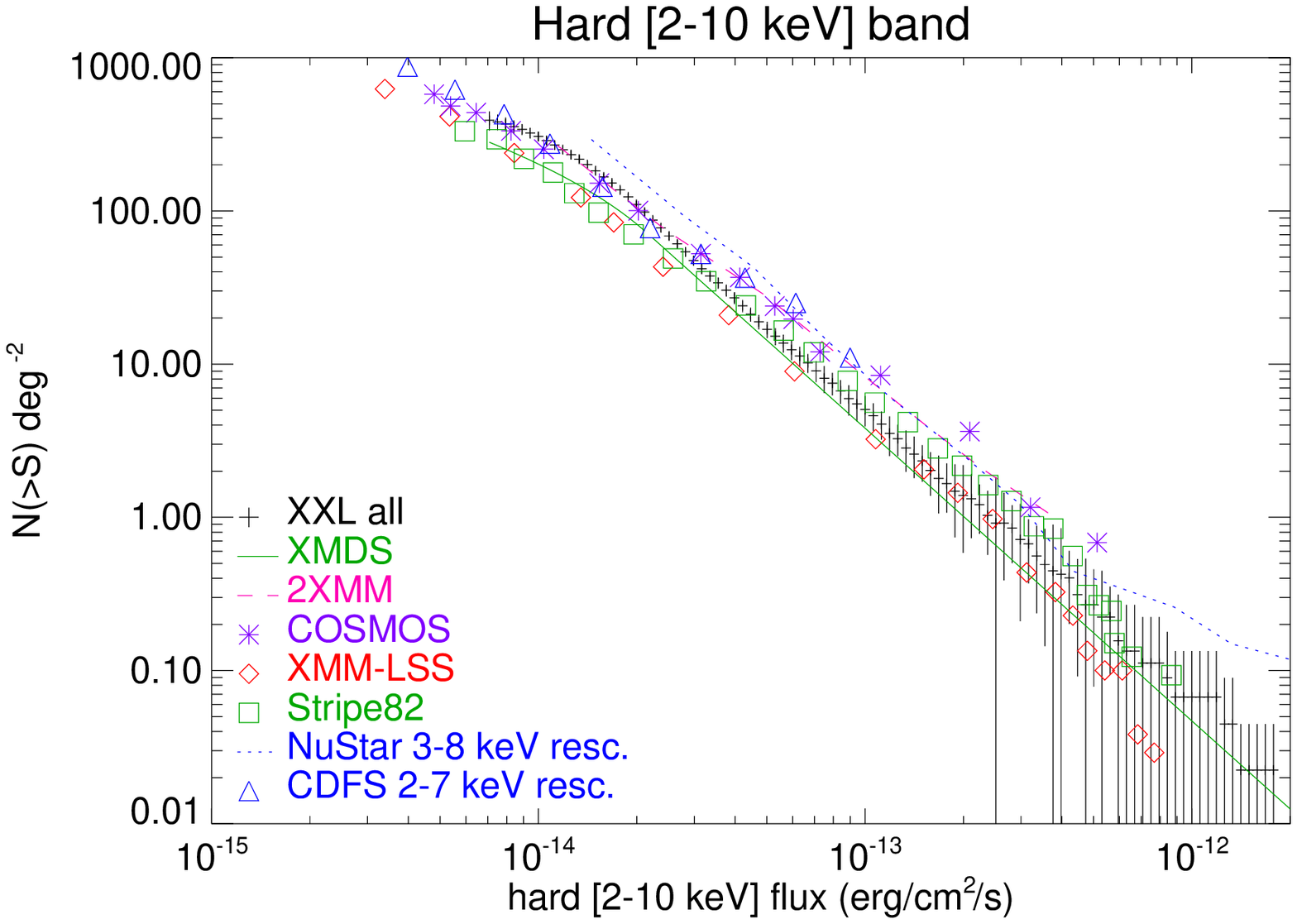}
 \caption{ $logN-logS$ relation for both XXL areas in the soft and hard X-ray bands
          (black crosses with $1\sigma$ error bars), compared with some literature
          measurements (see text for details).
  }
 \label{FigLogN}
\end{figure}

The $logN-logS$ relation has been
calculated (in the two energy bands) as described in \citet{2012A&A...537A.131E},
taking into account the numerically calculated probabilities to detect
sources with a certain flux, an off-axis distance in a pointing with
effective exposure, and a particle background level.
The sources in the overlaps between pointings were taken into account according to the Voronoi tessellation technique described in detail in
\citet{2012A&A...537A.131E}, and only good pointings with exposures below 15 ks were considered in order to have uniformity of coverage.
It is shown in Fig.~\ref{FigLogN} for the entire survey; the curves calculated separately for the XXL-N and XXL-S areas are extremely similar,
which indicates no effect of cosmic variance.

We  plot representative data from the literature.
We show the XMM-LSS data (which are a subset of XXL-N analysed with a previous version of
our pipeline, compare Table~\ref{TabCat} and Appendix~\ref{SecCompLSS}) reported by \citet{2012A&A...537A.131E},
plotted as red diamonds. We also plot  earlier data shown in the latter paper: a green solid line for the fit from
XMDS \citep[a subset of XMM-LSS analysed with a different pipeline,][]{2005A&A...439..413C},
and results from two independent surveys,
COSMOS \citep[violet star data points,][]{2007ApJS..172..341C} and
2XMM \citep[magenta dashed line,][]{2009A&A...500..749E}.
Among the more recent publications we show data from
\textit{NuSTAR} \citep[blue dotted line, for the hard band only, rescaled from 3-8 keV,][]{2016ApJ...831..185H},
Stripe82 \citep[observed by XMM, green squares,][]{2016ApJ...817..172L}, and
\textit{Chandra} CDFS 7Ms \citep[blue triangles,][]{2017ApJS..228....2L}. The hard-band points for CDFS were rescaled\footnote{
Band rescaling entails shifting the flux to the right, multiplying
by a factor equal to the ratio of the flux in the 2-10 keV band
to the flux in the original band, computed for an unabsorbed power law with $\Gamma=1.7$, consistently with the CFs in
Table~\ref{TabCF}.
}
from 2-7 keV.
Our $logN-logS$ is consistent with the result of earlier surveys, with COSMOS being the most deviant.
The slight difference with \citet{2012A&A...537A.131E} in the hard band could be due to the different pipeline
version or to the lower average exposure of the entire XXL versus the XMM-LSS pointings.

\section{Multiwavelength material \label{SecMW}}         

For the purpose of 
identifying the X-ray detections at other wavelengths,
we use optical, near-IR (NIR), IR, and UV 
photometric
data as described in Tables 3 and 4 of
\citetalias{2016A&A...592A...5F},
to which the reader is referred in particular for details about the limiting magnitudes
in the various bands, and for the data archives used.
In total, there are
eight surveys covering XXL-N and six surveys covering XXL-S in various filters, as
summarised in Table~\ref{TabSurv}
and in Sect.~\ref{SecPhotoSumm}.

Rather than using publicly available catalogues, we performed our own homogeneous
photometric extraction per filter and per tile, handling tile overlaps. All
magnitudes for all surveys 
are
in the AB system,
corrected to 
3\arcsec~ aperture magnitudes, and
corrected for galactic extinction
as described in \citetalias{2016A&A...592A...5F}.

The 
full
photometric catalogues, 
including
the photometric redshift computed as
described in \citetalias{2016A&A...592A...5F},
and all 
sources 
(not just
the counterparts of the X-ray sources)
will be published in a forthcoming paper \citep{Sotiria2018}.

The association of published spectroscopic redshifts with our counterparts is described in section~\ref{SecOurSpe}, while
more detail about the new AAOmega spectroscopic observations for which we publish
the complete catalogue is given in section~\ref{SecLidman}.

\begin{table}
\caption{Photometric surveys and filters used for counterpart association.
The mnemonic for surveys (all lowercase) and filters (capitalisation as shown)
is used to name the columns listed in Table~\ref{TabMulti}.
The IRAC filter mnemonics `36' and `45' correspond to the $3.6\mu$m and $4.5\mu$m filters.
The CFHT filter labelled `y' is an i-band replacement filter.
} \label{TabSurv}
\setlength{\tabcolsep}{1pt}
\begin{tabular}{llll    }
   \hline
   {\bf North} \\
   \hline
Survey & sub-surveys      & mnemonic & filters \\
   \hline
   SDSS   & --               & sdss    & u g r i z   \\
   CFHT   & D1 W1 WA WB WC   & cfht    & u g r i y z \\
   VISTA  & VHS VIDEO VIKING & vista   & z Y J H K   \\
   UKIDSS & UDS DXS          & ukidss  & J H K       \\
   WIRcam & --               & wircam  & K           \\
   IRAC   & --               & irac    & 36 45       \\
   GALEX  & DIS MIS AIS GI   & galex   & fuv nuv     \\
   WISE   & --               & wise    & w1 w2 w3 w4 \\
   \hline
   {\bf South} \\
   \hline
Survey & sub-surveys      & mnemonic & filters \\
   \hline
   BCS    & --               & bcs     & g r i z     \\
   DECam  & --               & decam   & g r i z     \\
   VISTA  & VHS              & vista   & J H K       \\
   SSDF   & --               & irac    & 36 45       \\
   GALEX  & MIS AIS GI       & galex   & fuv nuv     \\
   WISE   & --               & wise    & w1 w2 w3 w4 \\
   \hline
\end{tabular}
\end{table}

\subsection{Summary of photometric surveys\label{SecPhotoSumm}}

The XXL-N area is covered by the following surveys:

\begin{itemize}
\item  
Most of XXL-N was covered by the CFHT Legacy Survey
 (CFHTLS)
 Wide1 (W1) and Deep1 (D1) fields in five filters ($u,g,r,i,z$);
 we used the T0007 data release
 \citep{2007AAS...210.6202V,2012yCat.2317....0H}.
 Observations were obtained with the 3.6m Canada-France-Hawaii Telescope (CFHT), using
 the MegaCam wide field optical imaging facility.
 The northernmost part was observed at CFHT in the $g,r,z$ bands
 (PI: M. Pierre)
 and the relevant fields are labelled WA, WB, WC in Table~\ref{TabSurv}.

\item 
 The Sloan Digital Sky Survey (SDSS) also provides shallower $u,g,r,i,z$
 observations in the whole XXL-N area. We used the DR10 data release
 \citep{2014ApJS..211...17A}.

\item
XXL-N has been observed with the WIRcam camera on CFHT in the $K_s$ band
 (2.2 $\mu$m), see
 \citet{2016A&A...590A.102M}.

\item 
The UKIRT Infrared Deep Sky Survey (UKIDSS,
 \citealt{2006MNRAS.372.1227D})
 has two fields (in the J,H,K bands) targeted on the XMM-LSS area, enclosed in XXL-N:
 the Ultra Deep Survey (UDS), and the Deep Extragalactic Survey (DXS).
 We used the DR10 data release.
\end{itemize}

The XXL-S area is covered by the following surveys:

\begin{itemize}
\item 
The Blanco Cosmology Survey (BCS), using the Mosaic2 imager at
 the Cerro Tololo Inter-American Observatory (CTIO) 4m Blanco telescope,
 observed a large southern area,
 giving full coverage of XXL-S in the $g,r,i,z$ bands
 \citep{2009ApJ...698.1221M, 2012ApJ...757...83D}.

\item 
The Dark Energy Camera (DECam,
 \citealt{2015AJ....150..150F}),
 also on the CTIO Blanco telescope, provided deeper $g,r,i,z$ observations
 of XXL-S
 \citep{2015JInst..10C6014D}.
\end{itemize}

The following surveys cover both areas:

\begin{itemize}
\item  
The Vista Hemisphere Survey (VHS,
 \citealt{2013Msngr.154...35M}),
 using the Visible and Infrared Survey Telescope for Astronomy (VISTA)
 covered XXL-N and XXL-S in the J,H,K bands. Additional coverage of XXL-N
 is supplied by the
 VISTA Kilo-degree Infrared Galaxy Survey (VIKING, \citealt{2013Msngr.154...32E}) and
 VISTA Deep Extragalactic Observations Survey (VIDEO, \citealt{2013Msngr.154...26J}).

\item 
The Spitzer satellite has observed the large
 Spitzer South Pole Telescope Deep Field (SSDF), with its Infrared Array
 Camera (IRAC) in channel 1, 3.6$\mu$m, and 2, 4.5$\mu$m.
 This field fully covers XXL-S. See
 \citet{2014ApJS..212...16A}.
 Similar observations (PI: M. Bremer) were acquired over XXL-N.

\item 
The Wide-field Infrared Survey Explorer (WISE) mission observed
 the whole sky in four mid-IR (MIR) bands: W1=3.4$\mu$m, W2=4.6$\mu$m, W3=12$\mu$m, W4=22$\mu$m
 \citep{2010AJ....140.1868W}.
 We used the ALLWISE data release.

\item 
The GALEX satellite surveyed the entire sky 
 \citep{2005ApJ...619L...7M}
 in two ultraviolet
 bands, 1344-1786 \AA~(far-UV, FUV) and 1771-2831 \AA~(near-UV, NUV),
 with various surveys:
 All-Sky Imaging Survey (AIS), Medium Imaging Survey (MIS),
 and guest investigator (GI) programmes  in both XXL areas, and
 the Deep Imaging Survey (DIS) in XXL-N only.

\end{itemize}

\subsection{Counterpart association\label{SecIdent}}

In order to associate 
X-ray sources with
potential counterparts 
at other wavelengths,
we first generated multiwavelength counterpart sets,
matching the individual primary photometric catalogues (one per survey per band: 28 in XXL-N
and 19 in XXL-S, see Table~\ref{TabSurv}) 
among them
(using a matching radius of 0.7\arcsec~ for the optical and IR catalogues where
the image PSF is small, and 2\arcsec~ for GALEX, IRAC, and WISE) 
as described in \citetalias{2016A&A...592A...5F}.
For the further association with X-ray sources, 
we also considered  the case of counterparts in 
one
single primary catalogue.

For the association of 
entire
counterpart sets with the X-ray sources
within a search radius of 10\arcsec ,
we use 
two estimators. In one case, we computed the simple probability of chance coincidence according to
\cite{1986MNRAS.218...31D}
(done purely as a cross-check and 
for
reference since it is what 
was used
for the \texttt{XLSS} and
\texttt{2XLSS} catalogues):
\begin{equation}\label{EqP}
  P = 1 - exp(-\pi~ n(<m)~  d^2 )
\end{equation}

In the other case, we used 
the more robust likelihood ratio estimator 
\citep{1992MNRAS.259..413S}, which
has been used for
other surveys,  e.g. \cite{2007ApJS..172..353B}:
\begin{equation}\label{EqLR}
  LR = (q(m) exp(-0.5 d^2/\sigma^2)/2\pi\sigma^2) / K n(m)
\end{equation}

In both formulae, $d$ is the distance
in arcsec
between the X-ray position and the candidate counterpart position, 
the positional error $\sigma$ is assumed for simplicity equal to 1\arcsec,
while 
the dependency on
the magnitude $m$ of the counterpart
is  either via the sky density (sources per square arcsec)
$n(<m)$
of objects brighter than $m$, or the 
sky
density per magnitude bin $n(m)$.
                       The term $q(m)$ in Eq.~\ref{EqLR} is defined as
\begin{equation}\label{EqQ}
  q(m) = g(m) - K n(m)
,\end{equation}
where $g(m)$ is the sky density of putative true counterparts, i.e. 
unique objects within a suitable radius (in our case 3\arcsec) of each X-ray source,
and K is the ratio $N_U/N_{tot}$ between the number of such putative counterparts
$N_U$ and the total number $N_{tot}$ of all (also non-unique) objects in the band
close to the X-ray sources within the same radius, i.e. inclusive of background objects.

We computed these estimators for each survey and band where a potential
counterpart is present, and assigned the best value (highest LR)
to the counterpart set. We then ordered the counterpart sets by 
decreasing LR, and also divided them into three broad groups: good, fair,
and bad according to 
$LR>0.25$, $0.05<LR<0.25$, $LR<0.05$
using P as a cross-check.  

We then assign a preliminary rank, rejecting most of the cases with bad scores.
A \textit{primary single counterpart} 
is
either a physical solitary association, 
a
single non-bad association, or exceptionally
the best of the rejects,
which is `recovered'.
When
several
candidates above the thresholds exist, the one
with best estimator is considered the \textit{primary counterpart}, and 
all
the others are \textit{secondaries}. If the estimator
ratio between the primary and the best secondary is above 10, the primary should be
definitely preferred and the secondaries 
are
just nominally included. Otherwise
the primary is only nominally better than the secondaries, but it is indicative
of an intrinsic ambiguity.

Finally, we tie everything together by comparing the ranks assigned by the
different methods, cleaning up ambiguous counterpart sets sharing a counterpart
in more 
than one survey,
demoting to secondaries the detection in 
just the
surveys with poorer position resolution (WISE, IRAC, GALEX) and promoting the
cases with counterparts in several  surveys.

We assign a final rank 
(numbers
provided in Table~\ref{TabMWstat}) so that each X-ray source has a single
\textit{primary counterpart}, and zero, one, or more \textit{secondaries}.
Rejected counterparts  will not be listed in the multiwavelength catalogues.
The identification rank is coded by a three-character string. There will be always \emph{one primary} counterpart with
rank between 0 and 1, and there may be one or more \emph{secondary} counterparts with rank above 2. The codes are as follows:
\begin{itemize}
\item \textbf{0.0} physically single counterpart 
within 10\arcsec~ of the X-ray position;
\item \textbf{0.1} logically single counterpart (all others were rejected);
\item \textbf{0.2} `recovered' (the single counterpart has bad estimators);
\item \textbf{0.4} `blank field' (no catalogued counterpart 
within 10\arcsec~ of the X-ray position;
\item \textbf{0.9} the primary counterpart is much better than all the secondaries (ranked 2.2);
\item \textbf{1.0} the primary counterpart is just marginally better than one of the secondaries;
\item \textbf{2.1} this secondary counterpart is just marginally worse than the corresponding rank 1.0 primary;
\item \textbf{2.2} any other secondary counterpart.
\end{itemize}
Ranks below 0.2 correspond to single primaries; a rank of 0.9 corresponds to a preferred primary which is better than all
its secondaries; a rank of 1 to a nominal primary which is only marginally better than one of the secondaries.

\subsection{Bright star clean-up\label{SecStar}}

We have associated our counterparts with 
sources in
the SIMBAD and NED databases, and are providing an identifier for
the relevant cases. We used 
this
information as one way of partially screening out stars,
which are not of interest for AGN studies.
We also used
the USNO A2.0 catalogue
to screen stars, as well as
our own spectroscopic redshifts,
described in section~\ref{SecOurSpe}.

We assign a star flag of 1 to the counterparts with  $z_{spec}<0.003$, 
i.e. stars confirmed spectroscopically,
a flag of 2 to objects 
that look like a star upon
visual inspection, and a flag of 3 when both conditions occur. Visual inspection of thumbnail
optical images (see section \ref{SecMWDP}) was done systematically for all counterparts
with a 
match to a source
in a known \textit{star} catalogue in SIMBAD or NED. 
We also added objects in other catalogues (like 2MASS) which were inspected 
visually and found to be star-like.
We also  systematically inspected the counterparts with a 
match
in USNO A2.0 with a magnitude
brighter than $R<14$ (usually all the unnamed objects, i.e. not otherwise in SIMBAD or NED, are stars
and the rest are galaxies).

Of the 14\,168
primary counterparts in XXL-N
we find 163 `spectroscopic' stars, 
and visually identify 177 more; 23 were flagged as stars by both methods.
In XXL-S, the corresponding numbers are 384, 174, and 7.
We note that this screening is not at all complete.

\begin{figure}
 \includegraphics[width=8.2cm]{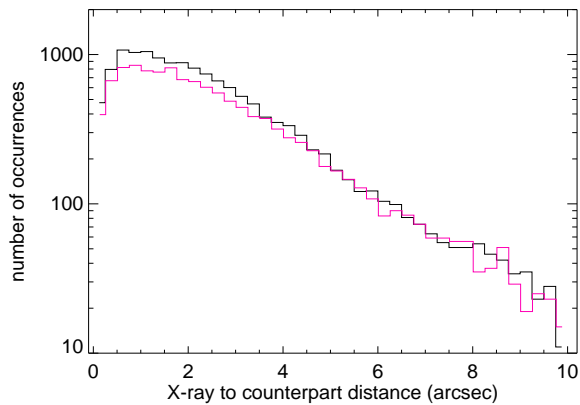}
 \caption{Distance 
 between the primary counterparts and the X-ray source for sources
 in XXL-N (in black) and  in XXL-S (in red). 
 }
 \label{FigDist}
\end{figure}

\subsection{Statistics\label{SecMWStat}}

Table \ref{TabMWstat} 
lists the number of counterparts sorted
by rank.
For 60\% of the cases (rank$<1$) there is a single 
counterpart or a counterpart that is clearly
preferred. In the remainder of the cases there is
some ambiguity in the identity of the real counterpart.

The number of X-ray sources 
that
have no apparent catalogued counterpart at all
(`blank fields') is very small: 18 in XXL-N and 21 in XXL-S. 
Not all blank fields are really empty (11 in XXL-N and 4 in XXL-S), 
as they might have been
affected by contamination
from
a nearby source which prevents
a detection, or the potential counterpart could actually be a bright
source 
that went undetected
by the photometric extraction.

To be conservative, one could exclude the rank of 0.2 counterparts (i.e. those 
where there is a single object 
near to
the X-ray source but
with bad association estimators).

\begin{table}
\caption{Basic statistics of the multiwavelength catalogues.
} \label{TabMWstat}
\begin{tabular}{llrr    }
   \hline
Rank & explanation      & XXL-N    & XXL-S   \\
   \hline
all  & distinct X-ray sources &  14\,168 & 11\,888 \\
   \hline
\multicolumn{4}{c}{Primary counterparts} \\
   \hline
0.0  & physically single    &         77 &      57 \\
0.1  & logically single     &     3\,462 &  2\,789 \\
0.2  & ``recovered'' bad    &        764 &     468 \\
0.4  & ``blank field''      &         18 &      21 \\
0.9  & definite primary     &     4\,138 &  3\,768 \\
1.0  & marginal primary     &     5\,709 &  4\,785 \\
   \hline
\multicolumn{4}{c}{Secondaries} \\
   \hline
2.1  & marginal secondary   &     5\,709 &  4\,785 \\
2.2  & any other secondary  &    11\,708 & 10\,850 \\
   \hline
   \hline
\end{tabular}
\end{table}

\begin{figure*}
 \includegraphics[width=16.4cm]{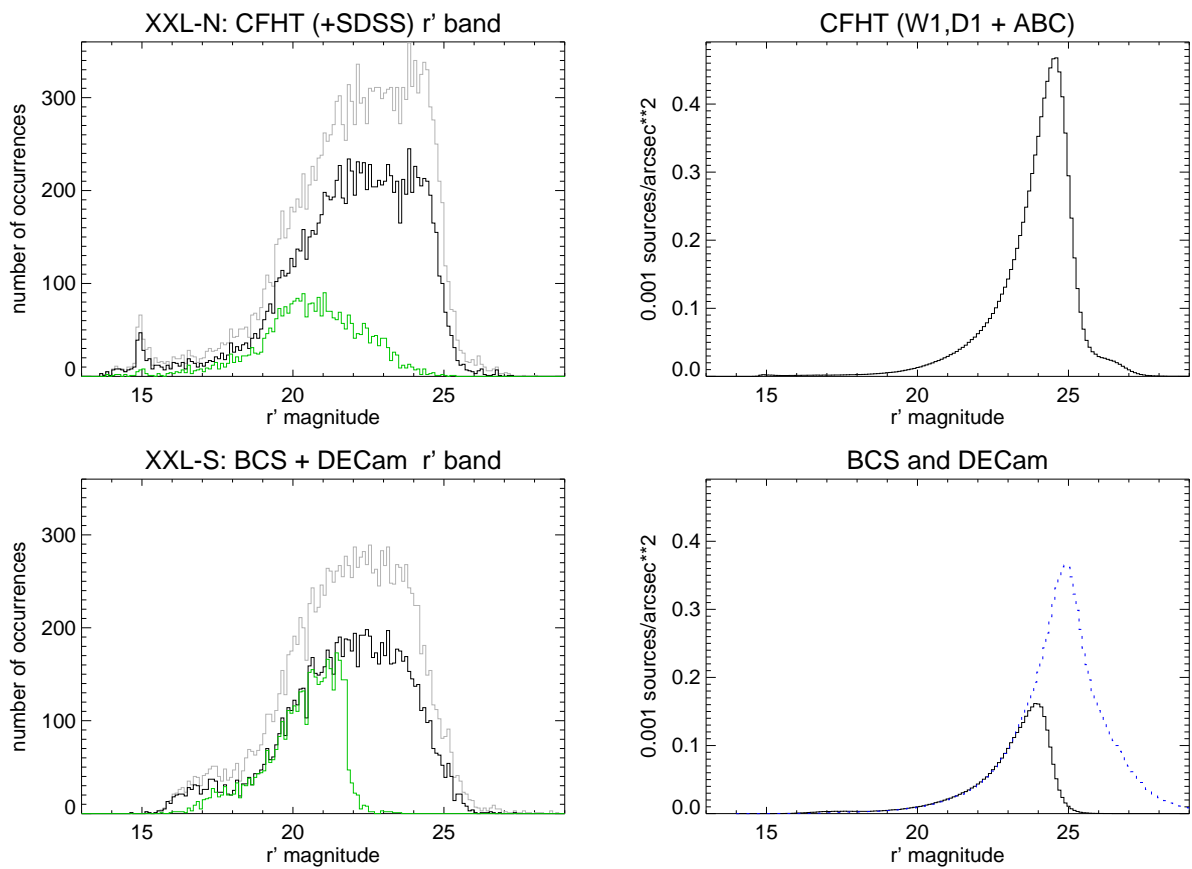}
 \caption{
 $r'$-band magnitude distribution for objects in
 XXL-N (top panels) and XXL-S (bottom panels).
          Left column: Histograms for counterparts:
           grey refers to all candidate counterparts (primaries and secondaries);
           black  to the primary counterparts; and  green to the subset of the
          primaries with a spectroscopic redshift.
          Right column: Densities $n(m)$
          for all objects in 
          specific
          surveys. For CFHT this amounts to more than 5 million 
          objects in $\sim$43 $deg^2$; for BCS (bottom panel, black solid curve) to approximately 2 million objects 
          in $\sim$47 $deg^2$; and for the deeper DECam survey (dashed blue curve) to approximately 2 million 
          in just $\sim$16 $deg^2$.
  }
 \label{FigMag}
\end{figure*}

\begin{figure*}
 \includegraphics[width=17.0cm]{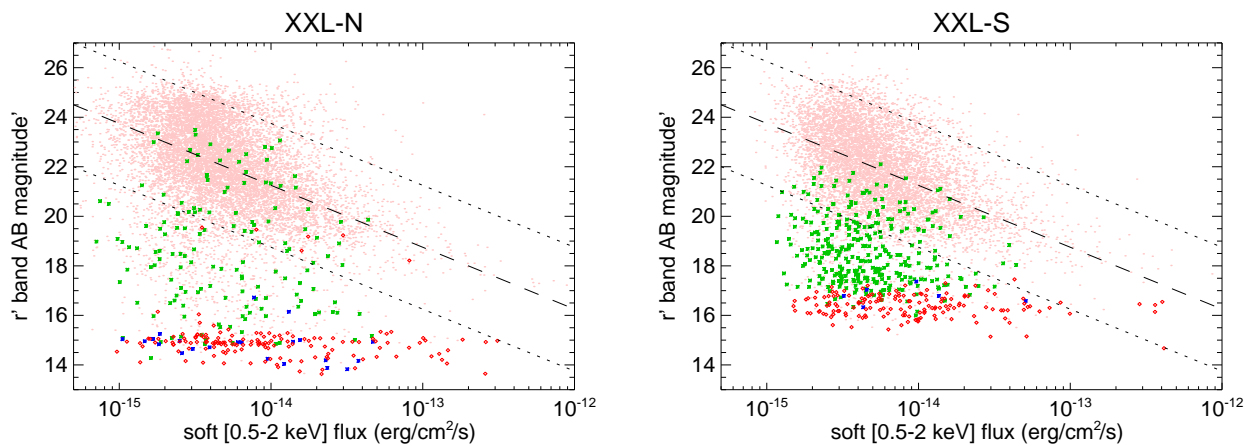}
 \caption{
Optical vs X-ray flux for XXL-N (left) and XXL-S (right).
  The $r'$ magnitude of the primary counterpart  is plotted against the broad-band soft flux [0.5-2 keV].
  Symbols and colour-codes are as follows:
  pink dots correspond to magnitudes available only from 
  either CFHT (preferred) or SDSS (in XXL-N), or
  either BCS (preferred) or DECam (in XXL-S);
  green asterisks to stars identified as such by their spectroscopic redshift (star flag 1);
  red diamonds to stars identified visually (star flag 2);
  blue asterisks to spectroscopic \textit{and} visual stars (star flag 3).
  The dashed and dotted lines mark the AGN locus of $X/O=0\pm1$ \citep[e.g.][]{2005ARA&A..43..827B}.
  }
 \label{FigMagFlux}
\end{figure*}

\begin{figure*}
 \includegraphics[width=17.0cm]{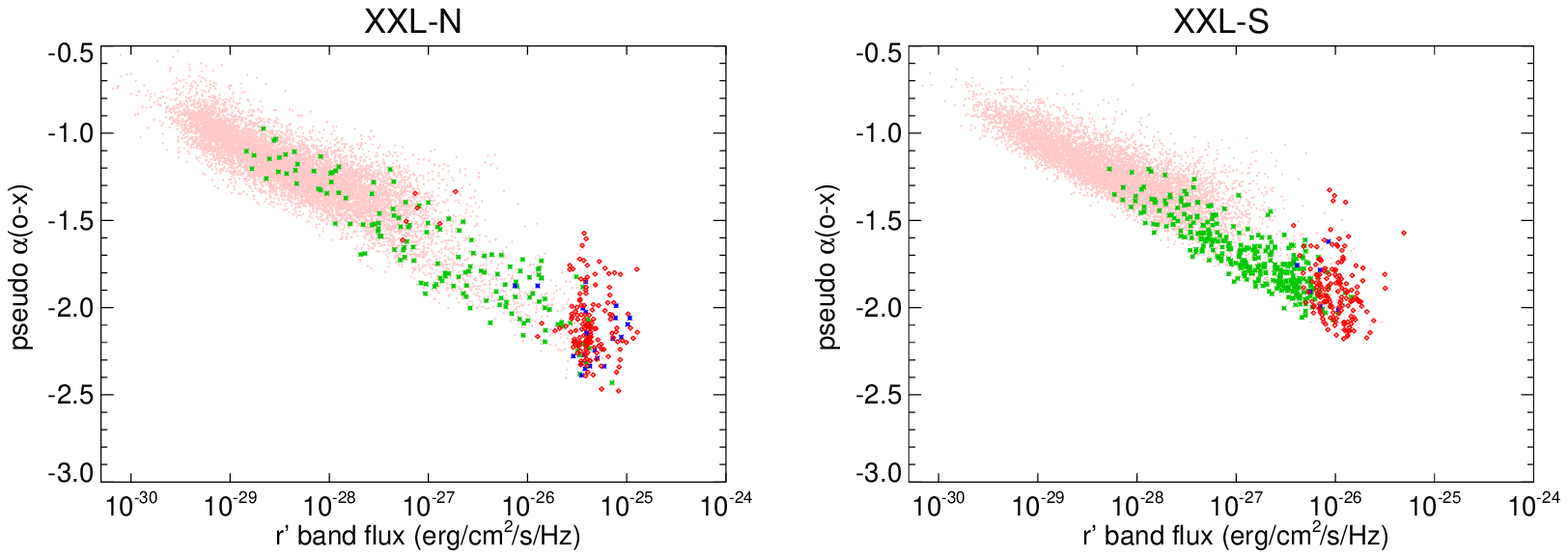}
 \caption{
Proxy for the $\alpha_{ox}$ index
for XXL-N (left panel) and XXL-S (right panel) as a function of optical flux.
The optical-to-X-ray spectral index is defined as $\alpha_{ox}=log(f_x/f_o)/log(\nu_x/\nu_o)$ where
the optical monochromatic flux is computed from the $r'$ AB magnitude $2.5log f_o = -r' -48.6$, and the reference
frequency $\nu_o$ derived from the $\lambda_{eff}$ of the filter (6258\AA~ for CFHT, 6185\AA~ for SDSS,
6266\AA~ for BCS, and 6433\AA~ for DECam); $\nu_x$ corresponds to 2 keV; the monochromatic 2 keV
flux $f_x$ is extrapolated from the observed [0.5-2 keV] soft flux $F_{soft}$ using the spectral model of Table~\ref{TabCF},
i.e. $f_x=7.626176\times10^{-27}*2*1.47115\times10^8 F_{soft}$.
Only sources detected in the soft X-ray band with a primary
counterpart in the $r'$ filter are considered.
  Symbols and colour-codes are as in Fig.~\ref{FigMagFlux}.
  }
 \label{FigAox}
\end{figure*}

In XXL-N about 67\% of the counterparts are observed in at least four surveys;
89\% of the primaries have a CFHT counterpart in some band; 77\% have IRAC;
70\% have a VISTA, WIRcam, or WISE detection; 39\% have GALEX; and 28\%
have 
a UKIDSS counterpart.

In XXL-S  about 61\% of the counterparts are observed in at least four surveys;
90\% of the primaries have a DECam counterpart; 80\% a BCS counterpart; 78\% have IRAC;
63\% VISTA; 44\% WISE; and 26\% 
have a GALEX counterpart.

For each X-ray source 
primary
counterpart, we compute the distance 
between
the astrometrically corrected X-ray position 
and
the counterpart in the survey 
that is nearest. A histogram of these distances is
shown in Fig.~\ref{FigDist}.

In XXL-N (XXL-S) 85\% (81\%)  of all primary counterparts 
are
within 4\arcsec.
This occurs for 94\% (93\%) of the best single primaries (ranks 0.0 and 0.1) or for
91\% (91\%) for the bona fide primaries (ranks 0.0, 0.1, and 0.9). Larger distances occur
for the marginal primaries (rank 1.0), with 74\% (69\%) within 4\arcsec, or for the `bad singles'
(rank 0.4) with 57\% (54\%).

In the left column of Fig.~\ref{FigMag} we 
plot,
as an example, the distribution of the magnitude of
the counterparts in the $r'$ filter (the other 
filters 
are similar) in XXL-N and
XXL-S. This is compared with the overall source density $n(m)$ (used in section~\ref{SecIdent})
from the corresponding surveys, 
shown
in the right column of the same figure.
We note that the spectroscopic coverage (green curves) has a sharper cut-off in XXL-S (and is more regular
under the cut-off), where the data is mainly based on pointed observations around our X-ray targets (see section~\ref{SecLidman}),
while in XXL-N the spectroscopic data comes from a variety of observations (see section~\ref{SecOurSpe}).

In order to compare the X-ray and optical fluxes, in Fig.~\ref{FigMagFlux} we plot the $r'$ magnitude versus the
soft (B) broad-band flux. In absence of a systematic X-ray spectral fit or SED analysis  (currently available only
for the 1000 brightest sources, see \citetalias{2016A&A...592A...5F}), we prefer to 
show
raw parameters.
The optical magnitude is taken from 
any available catalogue
(i.e. CFHT and SDSS in XXL-N, and BCS and DECam in XXL-S),
preferring CFHT or BCS when two measurements are present for a given counterpart. The different coverage characteristics
of the various surveys is apparent from the plots.
With reference to the star flag described in section~\ref{SecStar}, we see that the stars identified spectroscopically
(flag=1, green asterisks in 
the
figure) cover the entire space, while those identified visually (flag=2, red diamonds in 
the
figure)
are usually concentrated at the brighter magnitudes. In XXL-N there are six exceptions with $r'>18$: four of them are in
the PB faint blue star catalogue \citep{1984A&AS...58..565B}, one is in USNO A2.0, and one is LP 649-93. The last has a definite star-like appearance,
the others are unconspicuous, and two of them are reported by SIMBAD as possible quasars.
This gives an idea of the limited contamination of the method used in section~\ref{SecStar}.

An alternate view is presented in Fig.~\ref{FigAox},
where we compute a proxy for the optical-to-X-ray spectral index $\alpha_{ox}$,
using the broad-band [0.5-2] keV flux, instead of a monochromatic flux from a fit,
and the standard spectral model of Table~\ref{TabCF}.

\subsection{Associating redshifts with the counterparts \label{SecOurSpe}}

For the X-ray sources whose counterparts have spectroscopic observations, we report the redshift $z_{spec}$.
The redshifts may 
come
from large spectroscopic surveys with which we have collaborative agreements,
like VIPERS \citep{2014A&A...566A.108G} or GAMA 
\citep{2015MNRAS.452.2087L,2017arXiv171109139B}
in the XXL-N area; or, in XXL-S, from large campaigns of our own
(see \citetalias{2016PASA...33....1L} and section~\ref{SecLidman}); or from the compilation held in Marseille at
CESAM\footnote{\url{http://cesam.oamp.fr/xmm-lss/}}, which includes both published data of external
origin and smaller campaigns 
by
XXL PIs.
The origins are briefly listed in 
the
footnote to Table~\ref{TabMulti};
a full list with references is provided in Table~2 of  \citet{2017A&A...paperXXII},
hereafter \citetalias{2017A&A...paperXXII}. More details about the collection and reduction of spectroscopic
information at CESAM is provided in section~3 of \citet{2017A&A.....paperXX}, hereafter \citetalias{2017A&A.....paperXX}.
~
For the case where
the same counterpart has more than one redshift measurement, 
the procedure described in \citetalias{2017A&A...paperXXII} 
is adopted. Namely, it first
groups the originating
surveys by priority into three classes, and inside the same priority group
it takes
the spectrum with the best quality flag,
dividing  the survey-provided redshift
flags into four uniform classes.

In XXL-N we 
actually used
the very same sample used in \citetalias{2017A&A...paperXXII}, to which we re-added
the stars ($z_{spec}<0.003$) excluded there, and applied exactly the same quality choice.
For XXL-S we once more applied  an analogous recipe: the highest origin flag was assigned to the recent AAOmega
campaign (section~\ref{SecLidman}), then to the previous one from   \citetalias{2016PASA...33....1L}, then to the
other campaigns in the CESAM compilation, and finally to NED; the quality flags are identical to the ones in
\citetalias{2017A&A...paperXXII},
to which we refer for a complete description, except for
the `star' flag 6 listed in section \ref{sec:redshifts},
ignored in \citetalias{2017A&A...paperXXII}, but
considered equivalent to the `best' flag 4.

\begin{figure}
 \includegraphics[width=8.2cm]{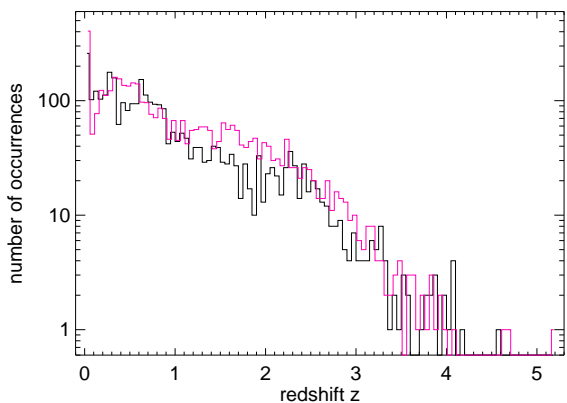}
 \caption{Redshifts of the primary counterparts in the XXL-N area (black) and
  in the XXL-S area (red). }
 \label{FigZ}
\end{figure}

Finally we associated the spectroscopic targets with our optical and NIR photometric potential counterparts 
within 1\arcsec. In XXL-N we dealt with a total of 5521 redshifts, of which 3235 refer to primary counterparts,
1098 to secondaries, and 1188 to rejected candidates. In XXL-S, excluding 540 sources observed with no valid redshift,
we have a total of 4745 redshifts (3873 primary counterparts, 661 secondaries and 211 rejects).
Figure~\ref{FigZ} gives the redshift distribution  of the primary counterparts.

In XXL-N 33\% of the redshifts of primary counterparts derive from GAMA, 30\% from VIPERS, 14\% from SDSS DR10, and
the rest from miscellaneous origins. In XXL-S 60\% of the redshifts come from \citetalias{2016PASA...33....1L},
38\% from our own 2016 observations described in the next section, and only a handful from other origins.

\subsection{The 2016 AAOmega spectra \label{SecLidman}}

In \citetalias{2016PASA...33....1L}, we noted that we 
used the AAOmega spectrograph \citep{2004SPIE.5492..410S} in
conjunction with the two-degree field (2dF) fibre positioner on the AAT
\citep{2002MNRAS.333..279L} to measure the redshifts of 3660 sources in the
XXL-S field. Here we report on new observations with the same
instrument.

\subsubsection{Target selection}

The targets were split into several categories,
as explained below. 
There were two categories for AGN,
one for radio galaxies, three for cluster galaxies, one for field
galaxies, and one for objects of special interest (e.g. gravitational
lenses and fossil groups).  The highest priority was for AGN, the next highest for
radio sources, followed by cluster galaxies and field galaxies.

The AGN were selected from our X-ray source list, targeting only objects brighter than $r_{AB}=21.8$.
Obvious bright stars were
removed by excluding objects that were 3\arcsec~ from a bright star in the UCAC4 catalogue.
There were two categories of AGN, those detected in the BCS survey, and all the rest.

The radio sources were selected from the ATCA 2.1 GHz survey of the XXL-S \citep{2017A&A..paperXVIII}, hereafter \citetalias{2017A&A..paperXVIII}.
Only radio sources brighter than $r_{AB} < 22.0$ (equivalent to $r_{vega}< 21.8$) were targeted, with a maximum of
120 radio sources per 2dF pointing. 

Candidate cluster galaxies were selected according to their position
and magnitude. For a given cluster, a galaxy was selected if it was
within 1\,Mpc of the cluster centre and had an r-band magnitude
within the range $m_{\star}-3.0 < r < m_{\star}+1.75$, where $m_{\star}$
is the r-band magnitude of an $L_{\star}$  galaxy at the redshift of the cluster.
Only cluster galaxy candidates brighter than $r_{AB}=22.5$ were targeted. There were three
categories of clusters. Clusters that lacked spectroscopic
confirmation had the highest priority.  
For these clusters, we used photometric redshift to estimate $m_{\star}$
and the angular distance corresponding to 1Mpc.
Next came clusters that were
spectroscopically confirmed, followed by clusters that were optically
identified.
We also assigned 17 fibres to candidate
fossil group members, as part of the search for fossil groups described
in \citetalias{2017A&A.....paperXX};
13 of them had redshifts with quality flags above 3, as defined in
section~\ref{sec:redshifts}. Of these, two were found to be stars.

Finally, randomly selected field galaxies brighter than $r_{AB}=21.8$ were
chosen to ensure that all fibres could be used.
Incidentally, we note that
one fibre was placed at the centre of a gravitational lens
independently rediscovered by one of the co-authors (GW) upon visual inspection
of DECam images, but already known as `the \object{Elliot Arc}' \citep{2011ApJ...742...48B}.
This is `\textit{a z = 0.9057 galaxy that is being strongly lensed by a massive galaxy cluster at a redshift of z=0.3838}'.
The lens is listed in our redshift catalogue as \object{XXL-AAOmega J235138.01-545254.2} at  z=0.379.
The cluster is \object{SCSO J235138-545253} \citep{2010ApJS..191..340M}.
Unfortunately, its position is in one of the outermost XXL pointings, XXLs053-08;
moreover, because it was at 15\arcmin~ from its centre,
it was not processed by \textsc{Xamin}, which stops at 13\arcmin, even though extended emission is visible in the X-ray image.

\subsubsection{Observations}

We observed 3572 targets in the XXL-S footprint during two
observing runs in 2016 (see Table~\ref{tab:observingDates}).

\begin{table}[h]
\caption{Observing dates}
\begin{center}
\begin{tabular}{ll}
\hline\hline
August 23 to August 28 & six half nights \\
September 24 to September 27 & four half nights\\
\hline\hline
\end{tabular}
\end{center}
\label{tab:observingDates}
\end{table}

The fields in XXL-S were first targeted with 2dF in 2013
\citepalias{2016PASA...33....1L}. On that occasion, the footprint of XXL-S was
covered with 13 2dF pointings. This time, we found a 
better optimised
solution that required one fewer pointing.

The set-up of AAOmega was almost identical to the set-up  used in
\citetalias{2016PASA...33....1L}, so we refer the reader to that paper for
details. We  adjusted the central wavelength of the blue grating
2\,nm further to the red (482\,nm). This allowed  more overlap
between the red and blue arms of AAOmega. The amount of overlap
between the two channels varies with fibre number, and is typically
5\,nm.  The spectra start and end at 370\,nm and 890\,nm,
respectively, with a spectral resolution of about 1500.

Science exposures lasted 40 minutes, and two of these were taken
consecutively in an observing sequence. Prior to each sequence, an arc
frame and two fibre flats were taken. The arc frame is used to calibrate
the wavelength scale of the spectra and the fibre flats are used to
determine the locations of the spectra on the CCD (called the
tramline map), to remove the relative wavelength dependent
transmission of the fibres (absolute normalisation is done using night
sky lines), and to determine the fibre profile for later use in
extracting the spectra from the science frames.

We also took dome flats. These were taken once for each plate at the
beginning of each of the runs, and were used in combination with the
fibre flats to determine the wavelength and fibre dependent transmission
of the fibres.

\begin{table}
\caption{ Number of objects targeted and the number of redshifts obtained
in the 2016 AAOmega campaign,
with a breakdown by quality flag. `Targeted' objects correspond to the sum of
all flags, and valid redshifts `obtained' to flags 3 to 6.
The total numbers in the last column also include  17 fossil group galaxies
and 1 lens galaxy, which are not listed in the breakdown in the previous columns.
}
\begin{center}
\begin{tabular}{rrrrrr}
\hline\hline
Quality &
AGN &
Radio &
Cluster &
Field &
Total \\
flag &
 &
Galaxies &
Galaxies &
Galaxies &
 \\
\hline
6        & 201  &  11  &     8 &  51        \\
4        & 891  & 749  &   305 &  63        \\
3        & 328  & 132  &    41 &  20        \\
2        & 178  &  83  &    42 &  22        \\
1        & 251  &  71  &    65 &  42        \\
\hline
Targeted & 1849 & 1046 &   461 & 198 & 3572 \\
Obtained & 1420 &  892 &   354 & 134 & 2813 \\
\hline\hline
\end{tabular}
\end{center}
\label{tab:redshiftSummary}
\end{table}

  \begin{table*}
   \caption{ XXL database tables of this (DR2) and the previous (DR1) releases.}
   \label{TabDR}
   \centering
   \begin{tabular}{llll}
    \hline\hline
    Table name              & CDS & Description    & Ref. \\
    \hline
    DR2 \\
    \texttt{xxlpointings}     & IX/52/xxlpoint & list of pointings and astrometric offsets        & 4 \\
    \texttt{3XLSS}            & IX/52/3xlss    & band merged X-ray catalogue   & 1 \\
    \texttt{3XLSSB}           & ---                  & soft band   X-ray catalogue    & 1 \\
    \texttt{3XLSSCD}          & ---                  & hard band   X-ray catalogue    & 1 \\
    \texttt{3XLSSOPTN}        & IX/52/3xlsoptn & multiwavelength   catalogue for XXL-N   & 2 \\
    \texttt{3XLSSOPTS}        & IX/52/3xlsopts & multiwavelength   catalogue for XXL-S   & 2 \\
    \texttt{XXL\_AAOmega\_16} & IX/52/xxlaaoz  & AAOmega redshifts          & 3 \\
    \texttt{XXL\_365\_GC}     & IX/52/xxl365gc & 365 brightest clusters     & 10\\
    \texttt{XXL\_ATCA\_16\_comp}    & IX/52/atcacomp   & final ATCA 2.1GHz radio catalogue (components) & 11 \\
    \texttt{XXL\_ATCA\_16\_src}     & IX/52/atcasrc    & final ATCA 2.1GHz radio catalogue (sources)    & 11 \\
    \texttt{XXL\_ATCA\_16\_ctpt}    & IX/52/atcactpt   & optical-NIR counterparts of ATCA 2.1GHz radio sources & 14 \\
    \texttt{XXL\_GMRT\_17}    & IX/52/xxl\_gmrt & GMRT 610MHz radio catalogue        & 12 \\
    \---                      & IX/52/xxlngal   & XXL-N spectrophotometric galaxy sample & 13 \\
    \---                      & IX/52/xxlnbcg   & XXL-N Bright Cluster Galaxies   sample & 15 \\
    \hline
    DR1 \\
    \texttt{xxlpointings}   & IX/49/xxlpoint    & list of pointings         & 4 \\
    \texttt{XXL\_100\_GC}   & IX/49/xxl100gc    & 100 brightest clusters    & 5 \\
    \texttt{XXL\_1000\_AGN} & IX/49/xxl1000a    & 1000 brightest pointlike  & 6 \\
    \texttt{XXL\_VLA\_15}   & IX/49/xxl\_vla    & VLA 3Ghz radio catalogue       & 7 \\
    \texttt{XXL\_ATCA\_15}  & IX/49/xxl\_atca   & pilot ATCA 2.1GHz radio catalogue    & 8 \\
    \texttt{XXL\_AAOmega\_15} & IX/49/xxlaaoz   & AAOmega redshifts         & 9 \\
    \hline
   \end{tabular}
   \tablebib{
    (1)~this paper, sect.~\ref{SecXrayTab};
    (2)~this paper, sect.~\ref{SecMWTab};
    (3)~this paper, sect.~\ref{SecLidman} (supplements \texttt{XXL\_AAOmega\_15});
    (4)~the table originally reported in \citetalias{2016A&A...592A...1P}
    has been incorporated in a new Table~\ref{TabAstrometry} in this paper adding astrometric offsets;
    (5)~\citetalias{2016A&A...592A...2P};
    (6)~\citetalias{2016A&A...592A...5F};
    (7)~\citet{2016A&A...592A...8B};    \citetalias{2016A&A...592A...8B};  
    (8)~\citet{2016A&A...592A..10S};    \citetalias{2016A&A...592A..10S};  
    (9)~\citetalias{2016PASA...33....1L};  
    (10) \citetalias{2017A&A.....paperXX}  (supersedes \texttt{XXL\_100\_GC});
    (11) \citetalias{2017A&A..paperXVIII} (supersedes \texttt{XXL\_ATCA\_15});
    (12) \citet{2017A&A...paperXXIX};  \citetalias{2017A&A...paperXXIX};
    (13) \citetalias{2017A&A...paperXXII};
    (14) \citet{2018A&A...paperXVI}; \citetalias{2018A&A...paperXVI};
    (15) \citet{2018A&A...paperXXVIII}; \citetalias{2018A&A...paperXXVIII};
   }
  \end{table*}

\subsubsection{Data processing}

After each run, we processed the raw data with a customised version of
6.32 of the {\tt 2dfdr}
pipeline\footnote{\url{http://www.aao.gov.au/science/instruments/AAOmega/reduction}}.
Further details on the processing of the data can be found in
\citetalias{2016PASA...33....1L} and 
\cite{2017MNRAS.472..273C}.

Once processed, we first spliced and then combined the data. Since some
targets
can appear in more than one field (because of field overlap) we summed
all the data on a single target into one spectrum, weighting on the
variance.

Redshifts were then measured (see Sect.~\ref{sec:redshifts}), and all
objects with secure redshifts were removed from the target catalogue
before the second run. This maximised the observing efficiency.

\subsubsection{Redshift estimates\label{sec:redshifts}}

We used {\sc marz} \citep{2016ascl.soft05001H} to inspect each spectrum and to
manually assign
a redshift. For each spectrum, we assign a quality flag that
varies from 1 to 6. The flags are identical to those used in the OzDES
redshift survey \citep{2015MNRAS.452.3047Y}, and have the following meanings:

\begin{itemize}
\item 6 - a star;
\item 4 - $> 99$\% probability that the redshift is correct;
\item 3 - $\sim 95$\% probability that the redshift is correct;
\item 2 - based on one or more very weak features;
\item 1 - unknown.
\end{itemize}

Only objects with flag 3, 4, and 6 are listed in our catalogue. Flag 5 is an
internal flag used by OzDES, and is not used here. The fraction of
objects with  flags set to 6, 4, or 3, or less than 3 were 0.08, 0.56, 0.15, and
0.21, respectively.

All redshifts are placed in the heliocentric reference frame.  The
typical redshift uncertainty for galaxies is  $0.0002 (1+z)$. For
AGN, it is $0.001 (1+z)$.

Of the 1849 objects targeted from the AGN catalogue, 1420 have
redshifts. The numbers for radio galaxies, cluster galaxies and field galaxies are
listed  in Table~\ref{tab:redshiftSummary}.


\section{The XXL database \label{SecDB}}

The catalogues and selected datasets of the XXL survey are publicly available in two ways:
\begin{itemize}
 \item Basic static catalogues are deposited at the CDS;
 \item Complete queryable catalogues are available via the XXL
       Master Catalogue browser\footnote{\url{http://cosmosdb.iasf-milano.inaf.it/XXL/}},
       in a database, which allows 
       one to select a subset
       of sources and
       to retrieve
 the       associated data products, and provides
       online documentation and help files.
\end{itemize}

The list of database tables associated with the earlier DR1 release or released with the
present paper or in other DR2 papers is given in Table~\ref{TabDR}.

\begin{table*}
\caption{ %
List of parameters provided in the public X-ray
catalogues. All are available at the XXL Milan database 
as
separate tables: {\tt 3XLSSB} for the soft band
and {\tt 3XLSSCD} for the hard band. The column name has a prefix when
there are two column names given (one with the prefix B and one
with the prefix CD). Only the column applicable to the given band
appears in the relevant table, but both may show up in the
band-merged table {\tt 3XLSS}.
Column names without prefixes are relevant to the
individual band only. The last four  columns indicate
respectively: (X)  whether a parameter is computed by
{\sc Xamin}, in which case the name given in Table 1 of
\citetalias{2017A&A...paperXXIV} is 
listed in the explanation column;
(m) whether a parameter is also available  in the
band-merged table; (o) whether a parameter is present in the multiwavelength
table together with those described in Table
\ref{TabMulti}; and (C) whether a parameter is present in the
catalogue stored at CDS. 
}\label{TabBand}
\setlength{\tabcolsep}{3pt}
\begin{tabular}{l|l|l|l|l|l|l}
   \hline
   Column name/s & units & explanation & X & m & o & C \\
   \hline\hline
{\tt Bseq}|{\tt CDseq}      & --    & internal sequence number (unique)                     &   & \checkmark & \checkmark & \checkmark \\
{\tt Bcatname}|{\tt CDcatname} & --    & IAU catalogue name {\tt 3XLSS{\it x} Jhhmmss.s-ddmmss}, {\tt{\it x}=B or CD} &   & \checkmark & \checkmark & \checkmark \\
{\tt Xseq}                & --    & numeric pointer to merged entry see Table \ref{TabMerge}      &   & \checkmark & \checkmark & \checkmark \\
{\tt Xcatname}            & --    & name pointer to merged entry see Table \ref{TabMerge}      &   & \checkmark & \checkmark & \checkmark \\
{\tt Xlsspointer}         & --    & {\tt Xseq} of corresponding source in \texttt{2XLSSd} catalogue &   & \checkmark & \checkmark & \checkmark \\
{\tt XFieldName}             & --    & XMM pointing name                                  &   & \checkmark & \checkmark &   \\
{\tt Xbadfield}           & 0|1|2|3 & pointing quality from best (0) to worst (3)         &   & \checkmark & \checkmark &   \\
{\tt expm1}               & s     & MOS1 camera exposure in the band                      & \checkmark &   &   &   \\
{\tt expm2}               & s     & MOS2 camera exposure in the band                      & \checkmark &   &   &   \\
{\tt exppn}               & s     & pn   camera exposure in the band                      & \checkmark &   &   &   \\
{\tt gapm1}               & \arcsec & MOS1 distance to nearest gap                        & \checkmark &   &   &   \\
{\tt gapm2}               & \arcsec & MOS2 distance to nearest gap                        & \checkmark &   &   &   \\
{\tt gappn}               & \arcsec & pn   distance to nearest gap                        & \checkmark &   &   &   \\
{\tt Bnearest}|{\tt CDnearest} & \arcsec & distance to nearest detected neighbour              & \checkmark & \checkmark &   &   \\
{\tt Bc1c2}|{\tt CDc1c2}  & 0|1|2   & 1 for class C1, 2 for C2, 0 for undefined                &   & \checkmark & \checkmark & \checkmark \\
{\tt Bp1}|{\tt CDp1}      & 0|1     & application of the P1 recipe\tablefootmark{a}            & & \checkmark & \checkmark & \checkmark \\
{\tt Bcorerad}|{\tt CDcorerad} & \arcsec & core radius \textsc{ext} (for extended sources)              & \checkmark & \checkmark &   & \checkmark \\
{\tt Bextlike}|{\tt CDextlike} & --      & extension statistic \textsc{ext\_stat}             & \checkmark & \checkmark &   & \checkmark \\
{\tt Bdetlik\_pnt}|{\tt CDdetlik\_pnt} & --      & detection statistic \textsc{pnt\_det\_stat} for point-like fit      & \checkmark &   &   &   \\
{\tt Bdetlik\_ext}|{\tt CDdetlik\_ext} & --      & detection statistic \textsc{ext\_det\_stat} for extended  fit      & \checkmark &   &   &   \\
{\tt Boffaxis}|{\tt CDoffaxis} & \arcmin & off-axis angle                                      &   & \checkmark &   & \checkmark \\
{\tt Brawra\_pnt}|{\tt CDrawra\_pnt}   & $\degr$ & source RA  (not astrometrically corrected) for point-like fit      & \checkmark &   &   &   \\
{\tt Brawdec\_pnt}|{\tt CDrawdec\_pnt} & $\degr$ & source Dec (not astrometrically corrected) for point-like fit      & \checkmark &   &   &   \\
{\tt Brawra\_ext}|{\tt CDrawra\_ext}   & $\degr$ & source RA  (not astrometrically corrected) for extended  fit      & \checkmark &   &   &   \\
{\tt Brawdec\_ext}|{\tt CDrawdec\_ext} & $\degr$ & source Dec (not astrometrically corrected) for extended  fit      & \checkmark &   &   &   \\
{\tt Bra\_pnt}|{\tt CDra\_pnt}         & $\degr$ & source RA  (astrometrically corrected) for point-like fit      &   &   &   &   \\
{\tt Bdec\_pnt}|{\tt CDdec\_pnt}       & $\degr$ & source Dec (astrometrically corrected) for point-like fit      &   &   &   &   \\
{\tt Bra\_ext}|{\tt CDra\_ext}         & $\degr$ & source RA  (astrometrically corrected) for extended  fit      &   &   &   &   \\
{\tt Bdec\_ext}|{\tt CDdec\_ext}       & $\degr$ & source Dec (astrometrically corrected) for extended  fit      &   &   &   &   \\
{\tt Bposerr}|{\tt CDposerr} & \arcsec & error on coordinates per Table 4 of \citet{2013MNRAS.429.1652C}   &   & \checkmark &   & \checkmark \\
{\tt Bratemos\_pnt}|{\tt CDratemos\_pnt} & ct/s/detector  & MOS count rate $CR_{MOS}$ for point-like fit         & \checkmark &   &   &   \\
{\tt Bratepn\_pnt}|{\tt CDratepn\_pnt}   & ct/s           & pn  count rate $CR_{PN}$ for point-like fit          & \checkmark &   &   &   \\
{\tt Bratemos\_ext}|{\tt CDratemos\_ext} & ct/s/detector  & MOS count rate $CR_{MOS}$ for extended fit           & \checkmark &   &   &   \\
{\tt Bratepn\_ext}|{\tt CDratepn\_ext}   & ct/s           & pn  count rate $CR_{PN}$ for extended fit           & \checkmark &   &   &   \\
{\tt countmos\_pnt  } & ct   & MOS number of counts for point-like fit          & \checkmark &   &   &   \\
{\tt countpn\_pnt   } & ct   & pn  number of counts for point-like fit          & \checkmark &   &   &   \\
{\tt countmos\_ext  } & ct   & MOS number of counts for extended fit           & \checkmark &   &   &   \\
{\tt countpn\_ext   } & ct   & pn  number of counts for extended fit           & \checkmark &   &   &   \\
{\tt bkgmos\_pnt    } & ct/pixel/detector & MOS local background for point-like fit          & \checkmark &   &   &   \\
{\tt bkgpn\_pnt     } & ct/pixel & pn  local background for point-like fit          & \checkmark &   &   &   \\
{\tt bkgmos\_ext    } & ct/pixel/detector & MOS local background for extended fit           & \checkmark &   &   &   \\
{\tt bkgpn\_ext     } & ct/pixel & pn  local background for extended fit           & \checkmark &   &   &   \\
{\tt Bflux}|{\tt CDflux}                & erg/cm$^{2}$/s & source flux (undefined i.e. -1 for extended)  &   & \checkmark &   & \checkmark \\
{\tt Bfluxerr}|{\tt CDfluxerr}          & erg/cm$^{2}$/s & error on source flux                          &   & \checkmark &   & \checkmark \\
{\tt Bfluxflag}|{\tt CDfluxflag}        & 0|1|2          & 0 if MOS-pn difference $<20\%$, 1 btw. 20\%-50\%, 2 $>50\%$ &   & \checkmark &   & \checkmark \\
   \hline
\end{tabular}
\tablefoot{\tablefoottext{a}{
\texttt{Bp1} or \texttt{CDp1}   is set to 1 if the nominal application of the P1 recipe (see section~\ref{SecPipe})
 is successful  in the soft or hard band, respectively. There is a limited number of sources (9 in XXL-N and 9 in XXL-S)  band-merged as
 \texttt{EP}, i.e. extended in the soft and pointlike in the hard band, for which the P1 recipe is successful in the hard
 band.
}
}
\end{table*}

\begin{table*}
\caption{List of database parameters,  as in Table  \ref{TabBand},
but for the additional columns present only in the merged
catalogue table {\tt 3XLSS}.  When there are two column names
given, one with the prefix B and one with the prefix CD, they
relate to the given band, and both show up in the band-merged
table. Column names with the prefix X are relevant to merged
properties. 
} \label{TabMerge}
\setlength{\tabcolsep}{3pt}
\begin{tabular}{l|l|l|l|l|l|l}
   \hline
   Column name/s & units & explanation &X & m & o & C \\
   \hline\hline
{\tt Xseq}                  & --      & Internal sequence number (unique)                 &   & \checkmark & \checkmark & \checkmark \\
{\tt Xcatname}              & --      & IAU catalogue name {\tt 3XLSS Jhhmmss.s-ddmmss{\it c}}, see section \ref{SecXrayTab}    &   & \checkmark & \checkmark & \checkmark \\
{\tt Bspurious}, {\tt CDspurious} & {\tt 1|0}  & set to 1 when soft/hard component has \textsc{pnt\_det\_stat$<$15} &   & \checkmark &   &   \\
{\tt Bdetlike}, {\tt CDdetlike}   & --      & detection statistic \textsc{det\_stat} (\textsc{pnt} or \textsc{ext} according to source class)           & \checkmark & \checkmark &   & \checkmark \\
{\tt Xra}                   & $\degr$ & source RA  (astrometr. corrected) (pnt or ext acc. to source class in best band) &   & \checkmark & \checkmark & \checkmark \\
{\tt Xdec}                  & $\degr$ & source Dec (astrometr. corrected) (pnt or ext acc. to source class in best band) &   & \checkmark & \checkmark & \checkmark \\
{\tt Bra}, {\tt CDra}           & $\degr$ & source RA  (astrometr. corrected) (pnt or ext according to source class) &   & \checkmark & \checkmark & \checkmark \\
{\tt Bdec}, {\tt CDdec}         & $\degr$ & source Dec (astrometr. corrected) (pnt or ext according to source class) &   & \checkmark & \checkmark & \checkmark \\
{\tt Xbestband}             & 2 or 3  & band with highest likelihood: 2 for B, 3 for CD  &   & \checkmark &   &   \\
{\tt Xastrocorr}            & 0|1|2|7 & astrometric correction from CFHTLS (7), BCS (1), USNO (2)  or  none (0) &   & \checkmark &   &   \\
{\tt Xmaxdist}              & \arcsec & distance between B and CD positions               &   & \checkmark &   &   \\
{\tt Xlink}                 & --      & pointer to Xseq of secondary association, see section \ref{SecXrayTab}  &   & \checkmark &   &   \\
{\tt Bratemos}, {\tt CDratemos}   & ct/s/detector   & MOS count rate $CR_{MOS}$ (pnt or ext according to source class)  & \checkmark & \checkmark &   & \checkmark \\
{\tt Bratepn}, {\tt CDratepn}     & ct/s   & pn  count rate $CR_{PN}$ (pnt or ext according to source class)  & \checkmark & \checkmark &   & \checkmark \\
{\tt Xextended}             & 0|1          & extended flag (based on \textsc{Xamin} C1/C2)                  & & \checkmark & \checkmark &  \\
{\tt XLSSC}                 & 1-1000       & XLSSC cluster number; 0 if not assigned, -1 if not applicable  & & \checkmark & \checkmark &  \\
{\tt agn1000}               & 1|0          & 1 if source listed in XXL Paper VI                             & & \checkmark & \checkmark &  \\
   \hline
\end{tabular}
\end{table*}

\begin{table*}
\caption{List of additional database parameters in the multiwavelength tables (in addition to the X-ray columns
 marked with a tick in column `o'
 of Tables \ref{TabBand} and \ref{TabMerge}).
} \label{TabMulti}
\begin{tabular}{l|l|l    }
   \hline
   Column name & units & explanation  \\
   \hline\hline
\multicolumn{3}{l}{Columns appearing once} \\
\hline
{\tt seq}                                 & --      & sequential counter of the counterpart set (kept internally but not provided) \\
{\tt Ctpra},{\tt Ctpdec}                  & degree  & counterpart coordinates from the closest survey \\
\hline
\multicolumn{3}{l}{Columns appearing once for each survey. Survey mnemonics {\it surv} are given in Table~\ref{TabSurv}}\\
\hline
{\it surv}{\tt seq}                       & --      & seq pointers to the various photometric tables (kept internally but not provided)\\
{\it surv}{\tt ra},{\it surv}{\tt dec}    & degree  & counterpart coordinates in the individual survey \\
{\it surv}{\tt dist}                      & arcsec  & X-ray to counterpart distance \\
\hline
\multicolumn{3}{l}{Columns appearing per survey and per band. Band (filter) mnemonics per survey are given in Table~\ref{TabSurv}}\\
\hline
{\it surv}{\tt mag}{\it band}             & AB mag  & 3\arcsec~aperture magnitude in given band for given survey \\
{\it surv}{\tt mag}{\it band}{\tt \_e}    & AB mag  & error on magnitude                       \\
\hline
\multicolumn{3}{l}{Again columns appearing once}\\
\hline
{\tt Xrank}                               &         &  Identification rank (see main text) \\
{\tt Pbest}                               & --      &  best band (smallest) chance probability (formula \ref{EqP})  \\
{\tt LRbest}                              & --      &  best band (highest) likelihood ratio (formula \ref{EqLR})    \\
\hline
{\tt zspec}                               & --      & spectroscopic redshift \\
{\tt origin}                              &         & origin of {\tt zspec}\tablefootmark{a} \\
\hline
{\tt simbadId}                            & --      & SIMBAD identifier (may be one of many) \\
{\tt nedId}                               & --      &    NED identifier (may be one of many) \\
{\tt star}                                & 0|1|2|3 & star flag: 1 if zspec=0, 2 by visual inspection, 3 both \\
{\tt agn1000}                             & -1|0|1  & 1 if counterpart in XXL Paper VI is confirmed,
                                                      -1 if X-ray source is listed in Paper VI  \\
                                          &         &  but counterpart changed,
                                                       0 if X-ray source not listed in Paper VI \\
   \hline
\end{tabular}
\tablefoot{          
\tablefoottext{a}{Spectroscopic surveys used are:
1: AAT\_AAOmega;                   
2: AAT\_AAOmega\_GAMA;              
3: Akiyama\_1;                     
4: Akiyama\_2;                     
5: Alpha\_compilation;             
6: ESOLP;                         
7: ESO\_FORS2\_600RI;               
8: ESO\_NTT;                       
9: AAOmega2012;
10: LDSS03;                       
11: Magellan;
12: Milano;                       
13: NED;                          
14: NED\_Vizier;                   
15: NED\_Vizier\_north;             
16: NED\_south;                    
17: NTT08;                        
18: NTT\_EFOSC2;                   
19: SDSS\_DR10;                    
20: SNLS;                         
21: Simpson2006;                  
22: Simpson2012;                  
23: Stalin;                       
24: Subaru;                       
25: VIPERS\_XXL;                   
26: VVDS\_UD;                      
27: VVDS\_deep;                    
28: WHT;                          
29: XMMLSS\_NTT;                   
30: XMMLSS\_PI\_FORS-LDSS;          
31: XMMLSS\_PI\_NTT05;              
32: XMMLSS\_PI\_NTT08;              
33: lidman15  \citepalias{2016PASA...33....1L};  
34: lidman16 (this paper, sect.~\ref{SecLidman}).
References for cases 1-32 are given in Table~2 of  \citetalias{2017A&A...paperXXII}.
}
}
\end{table*}

\subsection{X-ray tables\label{SecXrayTab}}

  In an analogous manner to
  the XMM-LSS releases \citep{2007MNRAS.382..279P,2013MNRAS.429.1652C}, we are
  providing a band-merged catalogue, called \texttt{3XLSS}, and single-band tables for the B [0.5-2] keV
  and CD [2-10] keV bands (called \texttt{3XLSSB} and \texttt{3XLSSCD}).
  Redundant sources detected in overlapping regions of different pointings are removed, as explained in
  section~\ref{SecPipe}. Only sources above a detection
statistic
  of 15 
(i.e. non-spurious)
  are made available in the 
  single-band tables, while the band-merged tables may include data with a likelihood below 15 in
  the other (non-best) band.

  For the table layout and column naming, we
  tried to be as consistent as possible with earlier XMM-LSS catalogues.
  The list of database columns 
  with brief descriptions
  is given in Tables \ref{TabBand} and \ref{TabMerge}. 
  Additional details about some columns are provided here:

  \begin{itemize}

  \item The official source catalogue name \texttt{Xcatname} is in the form
  \texttt{3XLSS Jhhmmss.s-ddmmss} to highlight the continuity with the IAU-registered names
  of the XMM-LSS catalogues (which the present catalogue overrides in the relevant sub-area).

  \item The single-band catalogue names \texttt{Bcatname} and \texttt{CDcatname} use the unofficial 
  prefixes \texttt{3XLSSB} or \texttt{3XLSSCD}.

  \item Band-merging ambiguities (see section~\ref{SecPipe}) may cause a detection
  in one band to be associated with two different objects in the other band. These 
  catalogue entries are flagged by a non-zero value in column \texttt{Xlink}, pointing to the
  \texttt{Xseq} of the other entry. 
  This might generate an
  ambiguity in the \texttt{Xcatname} 
  which
  is resolved
  (only four cases in XXL-N and six cases in XXL-S) by the addition of a suffix (e.g. the two
  members of a couple will appear as \texttt{3XLSS Jhhmmss.s-ddmmssa}
  and \texttt{3XLSS Jhhmmss.s-ddmmssb}).

  \item Column \texttt{Xlsspointer} assumes a non-zero value (only in XXL-N) if the source
  was already 
  listed in the \texttt{2XLSSd} catalogue 
  (taking the value of \texttt{Xseq} in \texttt{2XLSSd}).

  \item Column \texttt{Bc1c2} gives
  the C1/C2 classification for soft extended sources.
  The corresponding classification for sources nominally extended in the hard band is given in
  \texttt{CDc1c2}.
  A new \texttt{Xextended} flag tags all 
  sources flagged as
  extended by \textsc{Xamin}
  (see next subsection).

  \end{itemize}

  As we did for XMM-LSS, sources flagged as extended by \textsc{Xamin} are included in the catalogue,
  and now flagged 
  as such using
  column \texttt{Xextended}. This also includes  the nominal application of the C1/C2
  recipe to hard-only sources. 
  Their numbers are
  provided in Table~\ref{TabStat}.
  Their fluxes are set to -1 for all C1 sources, and left to the 
  flux computed according to Table~\ref{TabCF}
  for C2,
  consistent with the XMM-LSS usage (see caption of Table 9 in \citealt{2007MNRAS.382..279P}).
  An additional column \texttt{XLSSC} contains the number of the official designation \texttt{XLSSC nnn}
  of confirmed clusters (mostly published in \citetalias{2017A&A.....paperXX}). It contains 0 for
  candidates reserved for future analysis and -1 when not applicable (for pointlike sources).
  For the confirmed clusters we advise 
  looking for more meaningful parameters in the \texttt{XXL\_365\_GC}
  catalogue \citepalias{2017A&A.....paperXX}.

\subsection{X-ray data products\label{SecXrayDP}}

The per-pointing XMM photon and wavelet-smoothed images, and the camera exposure
maps were already released in \citetalias{2016A&A...592A...1P} as data products associated
with the pointing table \texttt{xxlpointings}. The same data products are also associated with the
\texttt{3XLSS} source catalogue table. With any query returning some X-ray sources (in some
pointings), it will be possible to retrieve the data products related to these pointings.

We note that the WCS of X-ray images is the one generated by \textsc{sas} and does not take into
account astrometric corrections. Strictly speaking one should use \emph{uncorrected} X-ray
positions to overlay on X-ray images and corrected positions to overlay on optical thumbnails 
(see section~\ref{SecMWDP}).
However, for X-ray images,
the difference is unimportant,
since the X-ray pixel size is 2.5\arcsec.

\subsection{Multiwavelength tables\label{SecMWTab}}

Since the number of multiwavelength catalogues available in the north and the south are different 
(summarised in Sect.~\ref{SecPhotoSumm}),
we supply two separate database tables for XXL-N and XXL-S, named 
3XLSSOPTN and 3XLSSOPTS.

The multiwavelength tables can have one or more records per X-ray source. They have one record
when there is no catalogued counterpart (blank fields) or exactly one counterpart, and more records for
ambiguous cases. 
Each record correspond to a multiwavelength counterpart set as defined in section~\ref{SecIdent}.
The X-ray information (see below) will be duplicated for all records
in our own database tables, and will be set to blank/undefined for secondary records at CDS.

The columns in the multiwavelength tables include a subset of the X-ray columns from the \texttt{3XLSS}
table 
(those with a tick in column `o' of Tables \ref{TabBand} and \ref{TabMerge})
plus the specific information in Table~\ref{TabMulti}:
the essential multiwavelength photometric information (position,
homogenised magnitudes, and errors), the 
spectroscopic redshift (see section~\ref{SecOurSpe}),
additional information like 
separations
or likelihoods, the rank of the counterpart
(primary/secondary), the identifier in SIMBAD or NED where available, and a
star flag.
There will be a multiplicity of database columns for the photometric information,
named as described in Table~\ref{TabMulti} according to the survey and band (filter)
mnemonics given in Table~\ref{TabSurv}: e.g. the RA coordinate from CFHT will be
named \texttt{cfhtra} and the one from GALEX \texttt{galexra}, the $z$ magnitude
from CFHT \texttt{cfhtmagz}, etc.

An AGN1000 flag 
applies to
the 1000 brightest sources already published in \citetalias{2016A&A...592A...5F}
(where X-ray spectral fits for the sources are available), and in particular if the counterpart matches.
We find that for 510 northern and 393 southern sources the primary counterpart matches the 
same counterpart chosen in the AGN1000 catalogue.
For 32 and respectively 44 cases the AGN1000 counterpart matches one of our secondaries, and for 5 and 3
cases it matches a rejected candidate.

\subsection{ AAOmega redshift catalogue}

The first four lines of the redshift catalogue are shown in
Table~\ref{tab:redshiftCatalogue} 
as 
an example
of the layout of the database table (analogous to previous version \texttt{XXL\_AAOmega\_15}).
The full catalogue can be obtained
from CDS,
via the XXL Master Catalogue Browser 
as table \texttt{XXL\_AAOmega\_16},
and from CESAM. 

\subsection{Multiwavelength data products\label{SecMWDP}}

For each X-ray source a number of thumbnail FITS images for the various surveys and       
bands are available as associated data products. 

The arrangement of the Master Catalogue database allows 
one to access
as data products the links to the
SIMBAD and NED pages for the sources which have such associations.

\begin{table}
\caption{Sample of the layout of the redshift catalogue}
\begin{center}
\begin{scriptsize}
\setlength{\tabcolsep}{1pt}
\begin{tabular}{llllc}
\hline\hline
Name & R.A. (J2000) & Dec. (J2000)& Redshift & Redshift flag \\
& (deg) & (deg) &  &  \\
\hline
XXL-AAOmega J231218.27-532406.0 & 348.07616 & -53.40167 &    0.7788  &  4\\ 
XXL-AAOmega J231218.67-533715.3 & 348.07782 & -53.62094 &    2.2062  &  4\\
XXL-AAOmega J231221.27-534529.1 & 348.08865 & -53.75811 &    0.0001  &  6\\ 
XXL-AAOmega J231233.28-532314.0 & 348.13867 & -53.38723 &    2.6824  &  4\\ 
\hline\hline
\end{tabular}
\end{scriptsize}
\end{center}
\label{tab:redshiftCatalogue}
\end{table}

\section{Concluding remarks\label{SecEnd}}

   In this paper, we have presented the intermediate release of the XXL
   X-ray catalogue (\texttt{3XLSS}) with 26\,056 sources over
   two 25 deg$^2$ areas. We have also included a multiwavelength catalogue
   with candidate counterparts
   in the UV, optical, NIR, and IR bands, and published the redshifts
   obtained with the AAOmega spectrograph in 2016 in the southern area.

   Catalogues and associated data products are available through the
   XXL Master Catalogue browser\footnote{\url{http://cosmosdb.iasf-milano.inaf.it/XXL/}}
   with a reduced summary stored at the CDS.

   A final version of the XXL catalogue is planned for 
   the end of 2018,
   with the reprocessing of
   all XXL data with the new \texttt{XAminF18} pipeline \citepalias{2017A&A...paperXXIV}.
   By combining events from overlapping pointings, we will make use of the full
   survey depth.

\begin{acknowledgements}
XXL is an international project based around an XMM
Very Large Programme surveying two 25 deg$^2$ extragalactic fields at a
depth of \textasciitilde 5 $10^{-15}$ erg s$^{-1}$ cm$^{-2}$ in [0.5-2] keV. The XXL website is
\url{http://irfu.cea.fr/xxl}. Multiband information and spectroscopic follow-up of the
X-ray sources are obtained through a number of survey programmes, summarised at
\url{http://xxlmultiwave.pbworks.com/}.
Based in part on data acquired through the Australian Astronomical Observatory, under programmes A/2016B/107.
This research has made use of the SIMBAD database, operated at CDS, Strasbourg, France, 
and of the NASA/IPAC Extragalactic Database (NED), 
which is operated by the Jet Propulsion Laboratory, California Institute of Technology, 
under contract with the National Aeronautics and Space Administration.
The Saclay team acknowledges long-term financial support from the 
Centre National d'Etudes Spatiales.
This paper uses data from the Galaxy And Mass Assembly (GAMA) survey and from the
VIMOS Public Extragalactic Redshift Survey (VIPERS). 
GAMA is a joint European-Australasian project based around a spectroscopic 
campaign using the Anglo-Australian Telescope. The GAMA input catalogue is 
based on data taken from the Sloan Digital Sky Survey and the UKIRT 
Infrared Deep Sky Survey. 
GAMA is funded by the STFC (UK), the ARC 
(Australia), the AAO, and the participating institutions. The GAMA website 
is \url{http://www.gama-survey.org/}.
VIPERS has been performed using the ESO Very Large Telescope, 
under the `Large Programme'182.A-0886. The participating institutions and 
funding agencies are listed at \url{http://vipers.inaf.it}.
The Saclay group acknowledges long-term support
from the Centre National d'Etudes Spatiales (CNES).
F.P. acknowledges support by the German Aerospace Agency (DLR) with funds
from the Ministry of Economy and Technology (BMWi) through grant 50 OR
1514 and grant 50 OR 1608.
S.A. acknowledges a post-doctoral fellowship from TUBITAK-BIDEB through 2219
programme.
E.K. thanks CNES and CNRS for support of post-doctoral research.
\end{acknowledgements}

%
%

\bibliographystyle{aa}
\bibliography{papercat}

\appendix
\section{Comparison with 
earlier 
XMM-LSS catalogues\label{SecCompLSS}}

\begin{figure}
 \includegraphics[width=8.2cm]{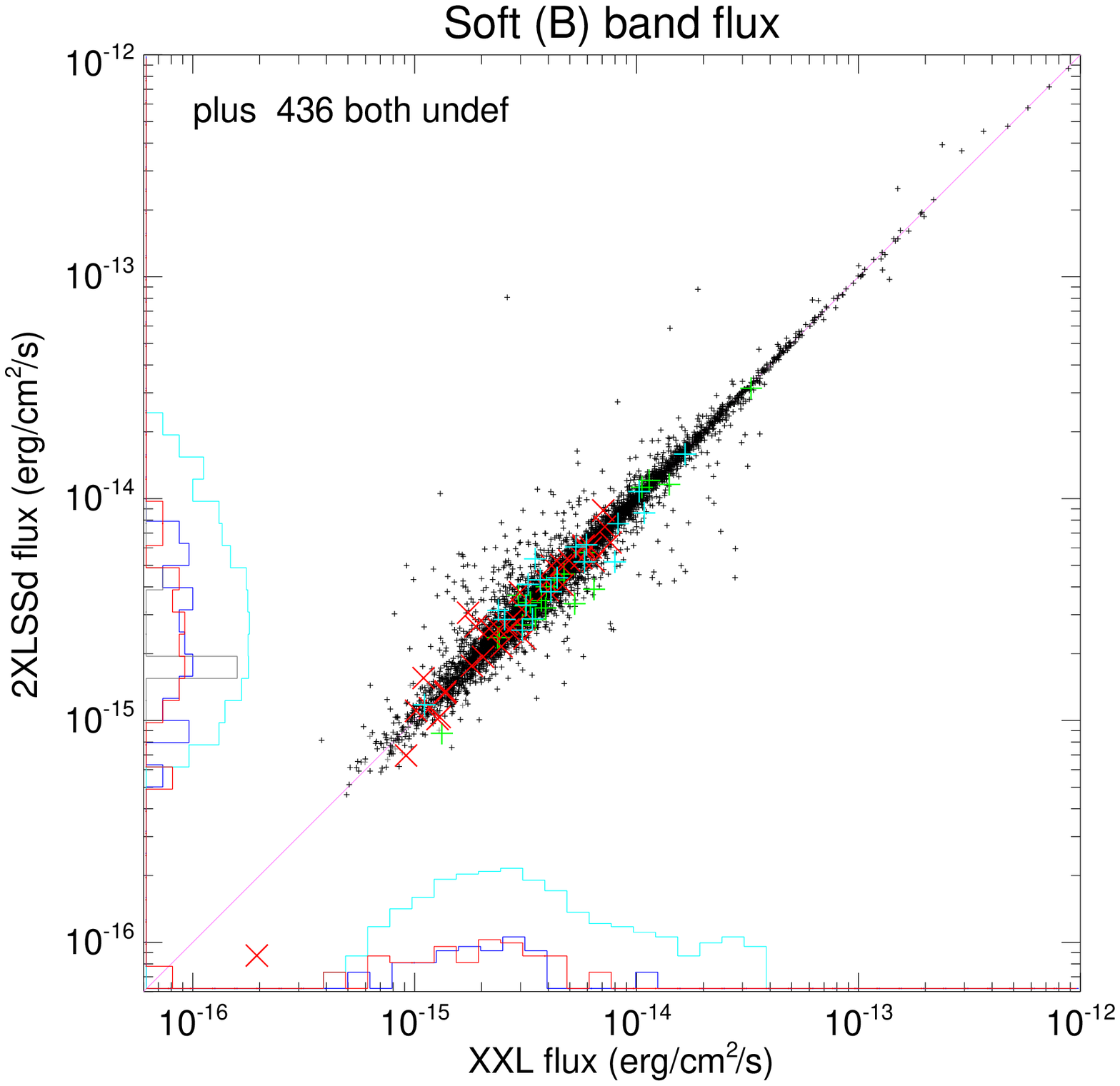}
 \includegraphics[width=8.2cm]{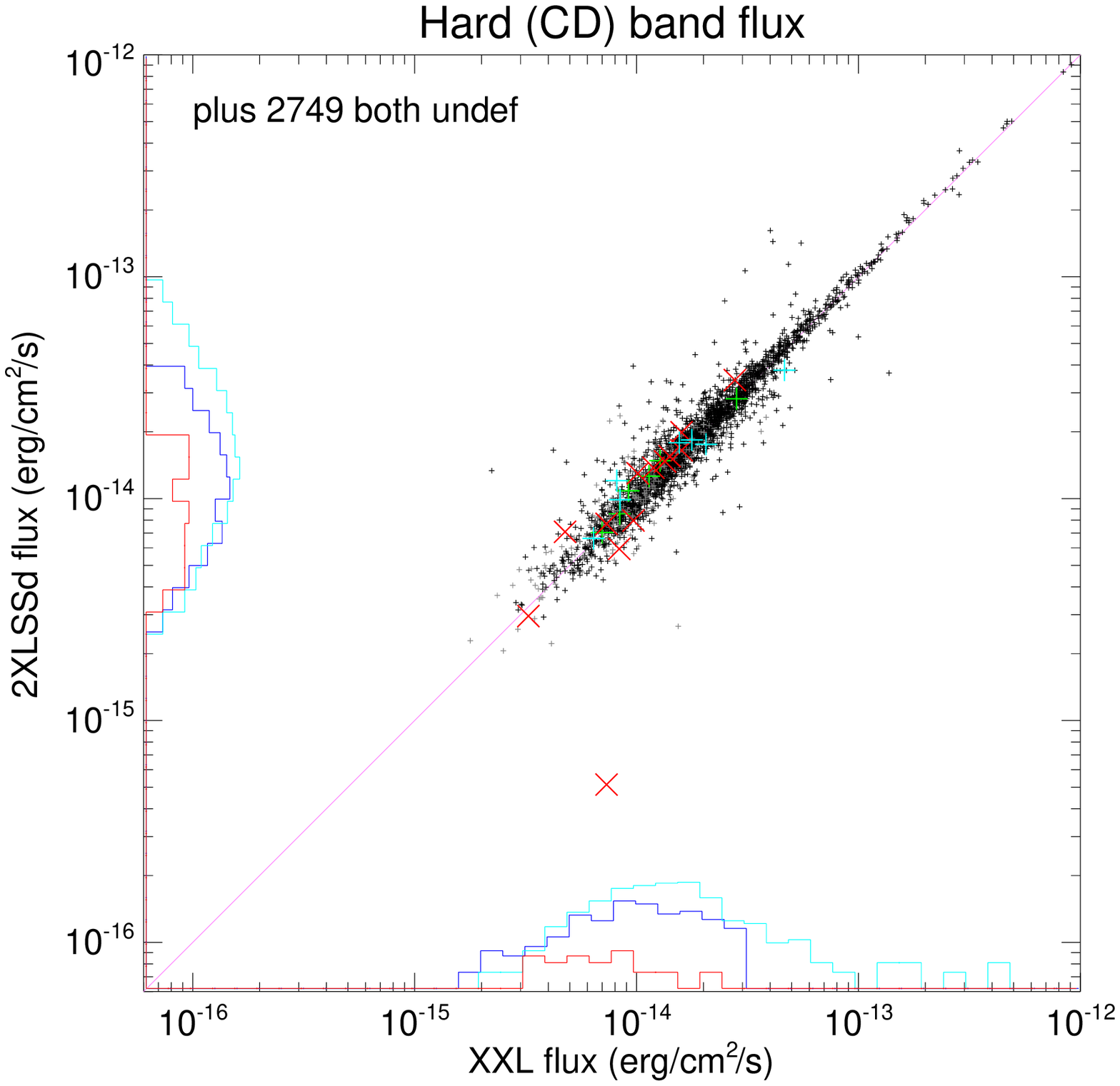}
 \caption{Comparison between XXL and \texttt{2XLSSd} fluxes in the soft (upper panel) and hard (lower panel) bands.
 Symbols and colour-coding are as follows: small black crosses are sources classified as pointlike in both the new and old
 catalogues; thick red Xs are sources classified as extended in both catalogues; large green crosses are  extended objects in the
 old catalogue and pointlike in the new one; cyan crosses are the reverse. The histograms
 (with arbitrary normalisations)
 indicate  sources present in both catalogues but having undefined
 flux in the energy band of interest in one catalogue (blue); extended sources detected only in one catalogue (red); and pointlike sources detected only in one catalogue (cyan). The extended sources consider only the C2 case, since C1 objects
 by construction have fluxes set to undefined in the database. The number of sources not detected in the band of interest
 in both catalogues is annotated in the top left corner.
 }
 \label{FigCompa}
\end{figure}

In order to assess the reproducibility across different versions of the \textsc{Xamin}
pipeline, we compare the present \texttt{3XLSS} catalogue with the XMM-LSS
\texttt{2XLSSd} \citep{2013MNRAS.429.1652C}, which overlaps a portion of XXL-N
(XMM-LSS pointings represent a subset fully enclosed in XXL; the same data were
reanalysed with a different version of the pipeline as indicated in Table~\ref{TabCat}).

There are 5827 entries that  have the column \texttt{Xlsspointer} (described in section~\ref{SecXrayTab}) set,
i.e. 
these sources
are in common between the old and new catalogue within a nominal 
separation
of 10\arcsec.
In only eight cases the same \texttt{Xlsspointer}
occurs twice, of which four correspond to ambiguous band merging (see section~\ref{SecBandMerge}).

\texttt{2XLSSd} included 6721 X-ray sources in 125 distinct XMM pointings. Of these, 902 sources
are not confirmed in the newer catalogue. Conversely, of the 7277 sources in the new catalogue
that are in the XMM-LSS pointings (those with identifiers of the form \texttt{XXLn000-{\it ppc}} or \texttt{XXLn998-{\it ppc}}),
1496 were not present in the older catalogue.
We note that \texttt{Xlsspointer} may not necessarily relate detections in the same pointing
because of differences in 
how the pointing overlap is treated,
although the pointing is the same in 86\% of the cases. In 3\% of the cases
a different XMM-LSS pointing was chosen; in 
the remaining 11\% 
one of the surrounding non-XMM-LSS
pointings was chosen.

Of the 902 
unconfirmed sources in \texttt{2XLSSd}, 91\% 
are single-band detections (only 78 are detected in both bands).
Most of them are very poor (64\% with detection 
statistics
 \textsc{det\_stat$<$20}),
only 7\% have \textsc{det\_stat$>$40} (which is about a $3\sigma$ detection
according to the cross-calibration in \citealt{2013MNRAS.429.1652C})
and less than 3\% \textsc{det\_stat$>$75} (about $4\sigma$).
Similarly, 94\% of the 1496 
newly detected sources
are single-band detections,
68\% are very poor (\textsc{det\_stat$<$20}), only 4\% above $3\sigma$,
and 2\% above $4\sigma$.
Conversely, of the 5827 common objects, 46\% are detected in both bands in one of the catalogues. Only
14\% of the \texttt{2XLSSd} and 11\% of the \texttt{3XLSS} 
sources
are below \textsc{det\_stat$<$20},
while about one-third in either catalogue are above \textsc{det\_stat$>$75}.
Not surprisingly, all the above means that the sources unaffected by the pipeline version
change are those with larger significance.

More than 99\% of the common objects have the same extended/pointlike flagging. Only 45
have changed classification, in equal proportion from pointlike to extended or vice versa;
the vast majority of the extended being classified C2.

Concerning the detection likelihood, this is usually comparable with a reasonable scatter.
It improves in the cases where \texttt{3XLSS} uses a different pointing,
possibly because one with better quality was chosen.

Also the fluxes, as shown in Fig.~\ref{FigCompa},
are usually comparable with a reasonable scatter (here the spread is
larger when \texttt{3XLSS} uses a different pointing).
The scatter can be quantified in terms of the percentage
deviation in absolute value between \texttt{2XLSSd} and \texttt{3XLSS} fluxes.
For all sources with a defined soft flux in both catalogues,   
66\% of the cases 
differ by
less than 10\%, 87\%                
are within 20\%, and 97\% within 50\%. In the hard band the 
percentages
are respectively
57\%, 82\%, and 97\%.                                         

\begin{figure}
 \includegraphics[width=8.2cm]{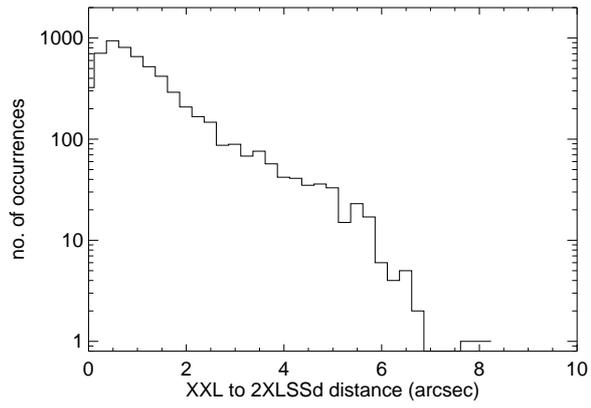}
 \caption{Distance
  between XXL and \texttt{2XLSSd} positions in astrometrically corrected coordinates for all common objects. }
 \label{FigDist2}
\end{figure}

Concerning the reproducibility of source positions, Fig.~\ref{FigDist2}
shows
the histogram of the
separation
between the position in the new and old catalogues, which takes into account both the 
differences in
pipeline version and the astrometric corrections.

\clearpage

\section{Pointing list and astrometric offsets\label{SecNewApp}}

Table~\ref{TabAstrometry} integrates 
the pointing list table, already published in \citetalias{2016A&A...592A...1P},
with the astrometric correction offsets described in section~\ref{SecAstroCorr}.

\begin{table*}
\caption{All XMM survey-type observations in the XXL fields,
integrated with astrometric correction offsets.
}
\begin{center}
\begin{small}
\setlength{\tabcolsep}{2pt}
\begin{tabular}{llrrrrrllllrrr}
\hline\hline
(1)   & (2)       & (3)& (4) & (5)  & (6)  & (7) & (8)    & (9)     & (10) & (11) & (12)     & (13)        & (14)\\
ObsId & FieldName & RA & Dec & MOS1 & MOS2 & pn & quality & badfield & db & cat & $\Delta$RA & $\Delta$Dec & corr\\
\hline
0037980101 & XXLn000-01a & 35.68970 & -3.84966 & 14.1 & 14.4 & 10.0 & 0 & 0 & X & X & $-1.072\pm0.428$ & $-1.072\pm0.395$ & CFHTLS \\
0037980201 & XXLn000-02a & 36.02333 & -3.85000 & 13.1 & 13.3 & 8.8  & 0 & 0 & X & X & $-1.072\pm0.383$ & $-0.528\pm0.381$ & CFHTLS \\
0037980301 & XXLn000-03a & 36.35712 & -3.84977 & 13.4 & 13.4 & 9.1  & 0 & 0 & X & X & $-0.544\pm0.421$ & $-0.528\pm0.422$ & CFHTLS \\
0037980401 & XXLn000-04a & 36.68933 & -3.85002 & 5.3  & 4.9  & 3.6  & 0 & 2 & X & X & $-0.528\pm0.543$ & $ 0.528\pm0.687$ & CFHTLS \\
0404960101 & XXLn000-04b & 36.64175 & -3.81891 & 8.9  & 9.0  & 3.3  & 0 & 2 & X & X & $-1.072\pm0.635$ & $-0.528\pm0.602$ & CFHTLS \\
0553910101 & XXLn000-04c & 36.64454 & -3.81438 & 11.2 & 11.5 & 8.6  & 0 & 2 & X & X & $-1.072\pm0.516$ & $-0.528\pm0.435$ & CFHTLS \\
0037980401 & XXLn000-04z & 36.64454 & -3.81891 & 25.3 & 25.4 & 15.5 & 0 & 0 & X & X & $-1.072\pm0.356$ & $-0.528\pm0.386$ & CFHTLS \\
0037980501 & XXLn000-05a & 37.02270 & -3.85013 & 15.9 & 15.9 & 11.8 & 0 & 0 & X & X & $ 0.528\pm0.340$ & $ 0.000\pm0.295$ & CFHTLS \\
0037980601 & XXLn000-06a & 35.52316 & -3.51672 & 13.0 & 13.0 & 8.8  & 0 & 0 & X & X & $-0.528\pm0.578$ & $-0.528\pm0.507$ & CFHTLS \\
0037980701 & XXLn000-07a & 35.85716 & -3.51575 & 12.3 & 12.3 & 7.8  & 0 & 0 & X & X & $-0.528\pm0.487$ & $ 0.528\pm0.486$ & CFHTLS \\
\hline\hline                                                                                                              
\end{tabular}                                                                                                             
\end{small}                                                                                                         
\tablefoot{                                                                                                               
 The full table is available at the CDS, and replaces the one published                                                   
 in \citetalias{2016A&A...592A...1P}.                                                                                     
 \texttt{FieldName} is the internal XXL labelling; $n (s)$ stands for the XXL-N (XXL-S)                                   
 field; $a,b,c...$ tags indicate that the same sky position has been observed                                             
 several times in different AOs (consult the ESA XMM log using the ESA ObsId)                                             
 because the quality of earlier pointings was insufficient; the z tag means that a                                        
 fictitious pointing has been created combining the events of all usable repeated                                         
 pointings in order to improve the quality.                                                                              
 In total there are 542 and 81 $a, b, c$, and z pointings, respectively.                                                   
 In case of repeated fields, and                                                                                          
 of overlaps from adjacent fields, the X-ray catalogue will remove overlapping                                            
 detections, and consider only the one from the better pointing, or, in case of                                           
 equal quality, the object with the smallest off-axis angle.                                                              
 Columns 5 to 7 give the remaining exposure (in ks) after selection of the                                                
 good-time intervals, for the MOS and pn detectors.                                                                       
 Quality Flag: 0 = Good quality,  1 = Low exposure,  2 = High background,  3                                           
 = 1 and 2. 
 Badfield Flag: 0 for best acceptable observation at a given position,                                                   
 1 for deep/good observation from the archives, not part of XXL proper,                                                  
 2 other acceptable XXL observation at same position,  3 bad pointings i.e.                                              
 quality=3. This flag is used in the overlap removal procedure.                                                           
 Column 10 is ticked if {\sc Xamin} detected at least one object in this pointing.                                        
 Column 11 is ticked if at least one source in this pointing survived the overlap                                         
 removal procedure and hence entered the X-ray source catalogue.                                                          
 Columns 12-13 (new in this version) report the astrometric offsets (in arcsec) computed                                  
 by \textsc{eposcorr} and Col. 14 indicates whether the correction was applied,                                         
 and from which reference catalogue (see section~\ref{SecAstroCorr} for details).                                         
}                                                                                                                         
\end{center}                                                                                                              
\label{TabAstrometry}                                                                                                     
\end{table*}

\end{document}